\documentclass[journal]{IEEEtran}
\usepackage[bottom]{footmisc}
\usepackage{amsmath,amssymb,amsfonts}
\usepackage{commath}
\usepackage{setspace}
\usepackage{cite}
\usepackage{graphicx}
\usepackage[center]{caption}
\usepackage{wrapfig}
\usepackage{hyperref}
\usepackage{pgfplots}
\pgfplotsset{width=10cm,compat=1.9}
\usepackage{tikz}
\usetikzlibrary{shapes.geometric}
\usepackage[ruled,vlined,linesnumbered]{algorithm2e}
\usepackage{textcomp}
\usepackage{xcolor}
\def\BibTeX{{\rm B\kern-.05em{\sc i\kern-.025em b}\kern-.08em
    T\kern-.1667em\lower.7ex\hbox{E}\kern-.125emX}}
\definecolor{light-blue}{RGB}{153, 204, 255}
\definecolor{eddA}{rgb}{0.82, 0.1, 0.26}
\definecolor{eddste}{rgb}{0.6, 0.4, 0.8}
\definecolor{ip}{rgb}{0.55, 0.71, 0.0}
\definecolor{greedy}{rgb}{0.0, 0.18, 0.39}
\definecolor{random}{rgb}{0.28, 0.02, 0.03}

\usepackage{scalerel}
\usepackage{tikz}
\usetikzlibrary{svg.path}

\definecolor{orcidlogocol}{HTML}{A6CE39}
\tikzset{
  orcidlogo/.pic={
    \fill[orcidlogocol] svg{M256,128c0,70.7-57.3,128-128,128C57.3,256,0,198.7,0,128C0,57.3,57.3,0,128,0C198.7,0,256,57.3,256,128z};
    \fill[white] svg{M86.3,186.2H70.9V79.1h15.4v48.4V186.2z}
                 svg{M108.9,79.1h41.6c39.6,0,57,28.3,57,53.6c0,27.5-21.5,53.6-56.8,53.6h-41.8V79.1z M124.3,172.4h24.5c34.9,0,42.9-26.5,42.9-39.7c0-21.5-13.7-39.7-43.7-39.7h-23.7V172.4z}
                 svg{M88.7,56.8c0,5.5-4.5,10.1-10.1,10.1c-5.6,0-10.1-4.6-10.1-10.1c0-5.6,4.5-10.1,10.1-10.1C84.2,46.7,88.7,51.3,88.7,56.8z};
  }
}
\newcommand\orcidicon[1]{\href{https://orcid.org/#1}{\mbox{\scalerel*{
\begin{tikzpicture}[yscale=-1,transform shape]
\pic{orcidlogo};
\end{tikzpicture}
}{|}}}}
\ifCLASSINFOpdf
\else
\fi
\begin{document}
\graphicspath{ {./Images/} }
\setstretch{0.9}
%

\title{Cost-Effective Edge Data Distribution with End-To-End Delay Guarantees in Edge Computing}

%
%

\author{Ravi Shankar\textsuperscript{ \orcidicon{0000-0003-4079-4979}}, Aryabartta Sahu\textsuperscript{ \orcidicon{0000-0002-5453-5022  }}~\IEEEmembership{IEEE Senior Member}, \ \\ Deptt. of CSE, IIT Guwahati, Assam, India \ \\ 
email:\{asahu,ravi170101053\}@iitg.ac.in}

%
%

\markboth{FUTURE GENERATION COMPUTER SYSTEMS}%
{Shell \MakeLowercase{\textit{et al.}}: Cost-Effective Edge Data Distribution with End-To-End Delay Guarantees in Edge Computing}
%



\maketitle

\begin{abstract}
Cloud Computing is the delivery of computing resources which includes servers, storage, databases, networking, software, analytics, and intelligence over the internet to offer faster innovation, flexible resources, and economies of scale. Since these computing resources are hosted centrally, the data transactions from the cloud to its users can get very expensive. Edge Computing plays a crucial role in minimizing these costs by shifting the data from the cloud to the edge servers located closer to the user's geographical location, thereby providing low-latency app-functionalities to the users of that area. However, the data transaction from the cloud to each of these edge servers can still be expensive both in time and cost. Thus, we need an application data distribution strategy that minimizes these penalities.\\

In this research, we attempt to formulate this Edge Data Distribution as a constrained optimization problem with end-to-end delay guarantees. We then provide an optimal approach to solve this problem using the Integer Programming (IP) technique. Since the IP approach has an exponential time complexity, we also then provide a modified implementation of the EDD-NSTE algorithm, for estimating solutions to large-scale EDD problems. These algorithms are then evaluated on standard real-world datasets named EUA and SLNDC and the result demonstrates that EDD-NSTE significantly outperformed, with a performance margin of 80.35\% over the other representative approaches in comparison.
\end{abstract}


%
\IEEEpeerreviewmaketitle

\section{Introduction}
\IEEEPARstart{C}{loud} Computing, in simplest form, is the practice of using a network of remote servers hosted on the internet to store, manage, and process data, rather than a local server or a personal computer \cite{IEEE-CC-ACM}. This facilitates the users to take advantage of the shared data center infrastructure and economy of scale to reduce costs. However, with an exponentially growing mobile data traffic promoted by a tremendous increase in mobile devices and IoT applications, the conventional centralized cloud computing paradigm \cite{rfc8793} \cite{Pathan2008} cannot handle this enormous increase in the network latency and congestion caused by these resources at the edge of the cloud. This is where Edge Computing kicks in.

Edge Computing can be traced back to the 1990s when Akamai launched its content delivery network (CDN), which introduced nodes at locations geographically closer to the end-users. These nodes stored cached static content such as images and videos. Edge Computing extends this concept by replacing these nodes with edge servers, thereby enabling computing and storage services at the edge of the cloud. This paradigm significantly improves the latency as it drastically reduces the volume of the data transmitted between servers and the end-users.

Since the edge servers improve the latency and bandwidth, the vendors of mobile and IoT applications (app vendors) can hire these computing and storage resources on the edge servers for hosting their applications. The users can then get rid of the computational burden and energy overload from their resource-limited end-devices to the edge servers located nearby. Thus, from an app vendors' perspective, caching data on these edge servers can considerably reduce both the latency for their users to fetch data and the volume of their app data transmitted between the cloud and its users, thereby reducing the transmission cost. 

Now, the app vendors' data transmission cost in the edge computing environment consists of two parts \textemdash \hspace{3pt}cloud-to-edge (C2E) and edge-to-edge (E2E) transmissions. The challenge is to formulate a strategy to distribute the application data onto these edge servers such that the app vendors' transmission cost is minimized, i.e., a cost-effective edge data distribution strategy is achieved.

Google AR \& VR\cite{GARVR} is an example of an industrial web application that benefits from incorporating an edge computing paradigm into its design framework. Augmented Reality (AR) \cite{GoogleARVR} expects the devices to process visual information and join pre-rendered visual components continuously. Without an edge computing design, this visual information needs to be conveyed back to the centralized cloud servers repeatedly which would add unavoidable critical latency, thereby reducing the application performance. It will also increase the total transmission cost, as the volume of the data transmitted will be huge. Thus, incorporating an edge computing framework is necessary to minimize these costs and improve the application's performance and user experience.

Now, once the edge computing framework is incorporated in the design and is set up in a geographical area with the construction of edge servers and the links between them, the company needs to prepare a strategy for caching the application data onto these edge servers. For example, it may decide to transmit the application data from the cloud to \textemdash
\begin{itemize}
	\item each of the edge servers directly or,
	\item some of the edge servers directly (initial transit edge servers), which will then distribute the data further to the other edge servers.
\end{itemize}

The challenge for the company here is to formulate an edge data distribution strategy that minimizes the total data transmission cost incurred within its latency constraint, i.e., Google's delay constraint must be fulfilled so that the transmissions do not take too long.

In this research, we attempted to study the edge data distribution problem considering the end-to-end delay guarantees, from the app vendors' perspective, to help them formulate an EDD strategy that minimizes the total EDD cost within the app vendor's latency constraint. The key contributions of this research are as follows:
\begin{itemize}
    \item We formulate and model the EDD problem \cite{IEEEexample:papermy} as a constrained optimization problem considering the total end-to-end delay, from the app vendor's perspective.
    \item We provide a refined formulation of an optimal approach to solve the Edge Data Distribution problem with end-to-end delay using the Integer Programming (IP) technique as compared to formulation given in Xia et al. \cite{IEEEexample:papermy}
    \item Since the IP approach has an exponential time complexity, thus, we also provide a modified version of an O(k) approximation algorithm named EDD-NSTE using network Steiner tree estimation (NSTE) for estimating solutions to dense large-scale EDD problems.
    \item The IP and EDD-NSTE solution approaches are then evaluated using standard EUA and SLNDC datasets and the results conclude that EDD-NSTE almost comparable to IP and outperformed the other three approaches in comparison namely \textemdash \hspace{3pt}Greedy, Random, and modified state-of-the-art EDD-A approach \cite{IEEEexample:papermy}.
\end{itemize}

The rest of the paper is arranged as follows: Section II reviews the previous and related works of this research. Section III describes the problem formulation of edge data distribution considering the total end-to-end delay guarantees. In Section IV we discuss the two solution strategies namely, with refined formulation using Integer Programming (IP) and proposed solution EDD-NSTE. Section V evaluates these algorithms along with other representative solution approaches and the modified state of art approach on the standard EUA and SLNDC datasets, and finally, we discuss the concluding remark of our work in Section VI.
\setlength{\arrayrulewidth}{1pt}
\begin{table}[t]
\caption{Summary of Common Notations} \label{tab:table1}
\setlength\tabcolsep{5pt}
\centering
\smallskip 
\begin{tabular*}{\columnwidth}{@{\extracolsep{\fill}}ll}
\hline
  \textbf{Notation} & \textbf{Description} \\ 
\hline
  $G$ & graph representing a particular region \\
  $N$ & number of edge servers in a particular region\\ 
  $V$ & set of all the edge servers in a particular region\\
  $E$ & set of all edges in the graph, i.e, transmission links \\
  $R$ & set of destination edge servers in the graph $G$\\
  $W$ & map of weights associated with each edge in $G$\\
  $u$, $v$ & edge servers \\
  $e\textsubscript{(u, v)}$ & edge between nodes $u$ and $v$\\
  $c$ & cloud server \\ 
  $S$ & set of binary variables indicating initial transit\\
      & edge server\\
  $H$ & set of binary variables indicating edge servers\\
      & visited during EDD process\\
  $T$ & set of binary variables indicating the data\\
      & distribution path\\
  $P\textsubscript{delay}$ & propagation delay caused by the transmission link\\
  $L\textsubscript{link}$ & length of the transmission link\\
  $s\textsubscript{medium}$ & speed of data transmission in the links\\
  $P\textsubscript{limit}$ & vendors' propagation delay limit\\
  $L\textsubscript{limit}$ & vendors' EDD length constraint, or depth limit\\
  $L\textsubscript{v}$ & depth of edge server $v$\\
  $\gamma$ & cost of any cloud-to-edge (C2E) transmission\\
  $\rho$ & destination edge server density \\
  $\delta$ & edge density \\
  $K$ & depth limit, $K = L\textsubscript{limit} - \gamma$\\
  $G\textsubscript{ST}$ & output graph of \hyperlink{algo:algo1}{Algorithm 1}\\
  $G\textsubscript{final}$ & output graph of \hyperlink{algo:algo3}{Algorithm 3}\\
  $G\textsubscript{DT}$ & directed graph with cloud as the root and edges\\
                        & replaced by the shortest path between two nodes\\
                        & in graph $G\textsubscript{ST}$\\
  $\overline{G}$ & metric closure of graph $G$\\
  $\overline{E}$ & set of edges in the metric closure of graph $G$\\
  $\overline{G}\textsubscript{R}$ & subgraph of $\overline{G}$ induced over $R$\\
  $mst(G)$ & minimum spanning tree of $G$\\
  $Triples$ & set of all combination of 3 destination\\
            & edge servers\\
  $G\textsubscript{cv}$ & tree having cloud as the root and edges\\
                        & $e \in G\textsubscript{final} \setminus \{c\}$\\
  $G\textsubscript{opt}$ & optimal solution of the given EDD problem\\
  $Cost(G)$ & cost associated with any graph $G$\\ 
  $v(z)$ & centroid of a particular $Triple$ $z$\\
  $d(e)$ & weight of edge $e$ in the graph\\
\hline
\end{tabular*}
\end{table}
\section{Related Work}
Cloud Computing is the delivery of computing resources which includes servers, storage, databases, networking, software, analytics, and intelligence over the internet to offer faster innovation, flexible resources, and economies of scale. Cloud or virtual storage service makes it possible to store data to a virtual or remote database and retrieve them anytime and from anywhere, upon user's request. The scheduling of these requests is a crucial problem in cloud computing and over the years, researchers have proposed a variety of scheduling algorithms. In a recent research \cite{IEEEexample:paperpiyush}, the authors suggested a state-of-the-art scheduler, which not only takes into consideration the load balance but also the application properties for request scheduling.

On the other hand, Edge Computing extends the Cloud Computing paradigm to the edge of the cloud by installing edge servers in locations geographically closer to the end-users. The placement of these edge servers in a given geographical region is a fundamental issue in an edge computing environment. In \cite{IEEEexample:paperA}, the authors suggested a cost-effective method that uses 1-0 integer programming to help providers of edge infrastructure make correct decisions regarding the edge servers placement. Similarly, Hao Yin et al.\cite{IEEEexample:paperB} suggested a decision support framework based on a flexible server placement, namely Tentacle, which aims to minimize the cost while maximizing the overall system performance.

The cost-effective application user allocation is another fundamental problem in edge computing first studied in \cite{IEEEexample:paperF}. In research \cite{IEEEexample:papersparsh}, the authors formulated it as a bin packing problem and suggested a heuristic approach for approximating the sub-optimal solutions to this problem.

In the recent researches \cite{IEEEexample:paperC}, \cite{IEEEexample:paperD}, \cite{IEEEExample:paperX}, \cite{IEEEExample:paperY}, \cite{IEEEExample:paperZ} and \cite{IEEEexample:paperE}, researchers have proposed investigative new techniques and approaches for data caching in the edge computing environment. However, even after having an optimal edge server placement, application user allocation, and data caching, we still need to consider the fact that transmitting the application data from the cloud to the edge servers also contributes a considerable amount to the app vendors' expense structure. Thus, a cost-effective application data distribution strategy is needed to minimize and accurately estimate the app vendors' total cost. A recent research \cite{IEEEexample:papermy} formulated this problem and suggested a heuristic state-of-the-art algorithm to estimate the total cost, including both the C2E and E2E transmissions.

However, existing research has not considered the fact that edge servers aren't located uniformly in terms of geographical locations, i.e, the delay to transmit the data between any two edge servers cannot always be the same.  Thus, instead of the hop distances as the vendor's latency constraint, we need to consider the total end-to-end delay as the EDD latency parameter. To the best of our knowledge, this paper makes the first attempt to solve the EDD problem considering the total end-to-end delay guarantees and introduces modified heuristic approaches to estimate the total cost associated with the given EDD from the app vendors' perspective.

\section{Problem Formulation}
To practice an edge computing paradigm in a specific geographical region, we first need to install an edge-server network in that area. The edge-server network is a physical network of edge-servers connected via transmission links (used for communicating and sharing data among the edge servers). This edge-server network can then be formulated as a graph, where nodes of the graph represent the edge servers, the edges represent the transmission links, and edge-weights represent the length of the links or, equivalently, the propagation delay for data transmission between these edge-servers. 

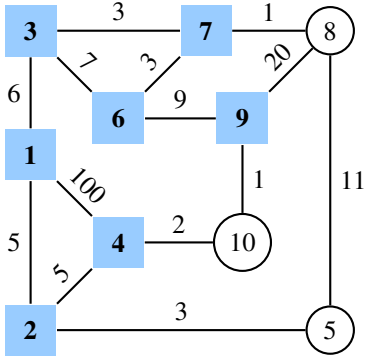
\begin{figure}[t]
\begin{center}
\begin{tikzpicture}[node distance={16.5mm}, thick, main/.style = {draw, circle}, square/.style={regular polygon, regular polygon sides=4}] 
\node[square] (6) [fill=light-blue] {\textbf{6}};
\node[square] (9) [fill=light-blue,right of=6] {\textbf{9}};
\node[main] (10) [below of=9] {10};
\node[square] (4) [fill=light-blue, below of=6] {\textbf{4}};
\node[square] (3) [fill=light-blue, above left of=6] {\textbf{3}};
\node[square] (7) [fill=light-blue, above right of=6] {\textbf{7}};
\node[square] (1) [fill=light-blue, above left of=4] {\textbf{1}};
\node[square] (2) [fill=light-blue, below left of=4] {\textbf{2}};
\node[main] (5) [below right of=10] {5};
\node[main] (8) [above right of=9] {8};

\draw[-] (1) -- (2) node[midway, left] {5}; 
\draw[-] (1) -- (3) node[midway, left] {6};
\draw[-] (1) -- (4) node[midway, above, sloped] {100};
\draw[-] (3) -- (6) node[midway, above, sloped] {7}; 
\draw[-] (6) -- (7) node[midway, above, sloped] {3};
\draw[-] (6) -- (9) node[midway, above] {9};
\draw[-] (4) -- (10) node[midway, above] {2}; 
\draw[-] (10) -- (9) node[midway, right] {1};
\draw[-] (2) -- (5) node[midway, above] {3};
\draw[-] (5) -- (8) node[midway, right] {11};
\draw[-] (9) -- (8) node[midway, above, sloped] {20};
\draw[-] (3) -- (7) node[midway, above] {3};
\draw[-] (7) -- (8) node[midway, above] {1};
\draw[-] (2) -- (4) node[midway, above, sloped] {5};
\end{tikzpicture}
\caption{EDD scenario with 10 edge servers} \label{fig:fig2}
\end{center}
\end{figure}

In this research, we consider an edge-server network of a particular geographical region and model it as a graph $G(V, E, W)$, with $V$ being the number of edge servers, $E$ as the set of edges or transmission links, and $W$ is the weights associated with each edge. Now, let $R$ denote the set of destination edge-servers in the graph $G(V, E, W)$, i.e., the edge servers which must be sent data from the cloud server within the vendor's specific length constraint. For example, in \autoref{fig:fig2}, $N$ = 10, which means that we have 10 edge servers installed in a geographical region corresponding to $V$ = \{1, 2, 3, 4, 5, 6, 7, 8, 9, 10\}. The set of edges $E$ = \{$e\textsubscript{(1, 4)}$, $e\textsubscript{(1, 2)}$, $e\textsubscript{(1, 3)}$ $\cdots$ $e\textsubscript{(9, 8)}$\} represents the transmission links. The weights corresponding to these edges are $W$ = \{100, 5, 6, $\cdots$ 20\} respectively. Similarly, the set of destination edge servers $R$ = \{3, 7, 6, 9, 1, 4, 2\}. The common notations used in this paper are listed in \autoref{tab:table1}.
\begin{figure*}[t]
\centering
\begin{minipage}[b]{0.48\textwidth}
\centering
    \begin{tikzpicture}[node distance={20mm}, thick, main/.style = {draw, circle},new/.style = {draw, circle, line width=1.5pt}, square/.style={regular polygon,regular polygon sides=4}] 
    \node[new] (1) [] {\textbf{1}};
    \node[new] (2) [right of=1] {2};
    \node[new] (3) [right of=2] {3};
    \node[square] (4) [fill=light-blue, below of=2] {4};
    \node[new] (5) [below of=1] {5};
    \node[main] (6) [below of=5] {6};
    \node[main] (7) [right of=6] {7};
    \node[new] (8) [below of=3] {8};
    \node[main] (9) [below of=8] {9};
    
    \draw[line width = 2pt, ->] (1) -- (2) node[midway, above] {2}; 
    \draw[line width = 2pt, ->] (2) -- (3) node[midway, above] {7};
    \draw[line width = 2pt, ->] (1) -- (5) node[midway, left] {9};
    \draw[-] (5) -- (6) node[midway, left] {12};
    \draw[-] (6) -- (7) node[midway, above] {3};
    \draw[line width = 2pt, <-] (1) -- (4) node[midway, above, sloped] {1};
    \draw[-] (4) -- (7) node[midway, right] {27};
    \draw[-] (7) -- (9) node[midway, above] {8};
    \draw[-] (3) -- (8) node[midway, right] {4};
    \draw[-] (8) -- (9) node[midway, right] {6};
    \draw[line width = 2pt, ->] (4) -- (8) node[midway, above] {10};
    \end{tikzpicture}
    \caption{EDD example to demonstrate L\textsubscript{limit}} \label{fig:fig3}
\end{minipage}
\begin{minipage}[b]{0.48\textwidth}
\centering
    \begin{tikzpicture}[node distance={16.5mm}, thick, main/.style = {draw, circle}, square/.style={regular polygon,regular polygon sides=4}] 
    \node[square] (6) [fill=light-blue] {\textbf{6}};
    \node[square] (9) [fill=light-blue,right of=6] {\textbf{9}};
    \node[main] (10) [below of=9] {10};
    \node[square] (4) [fill=light-blue, below of=6] {\textbf{4}};
    \node[square] (3) [fill=light-blue, above left of=6] {\textbf{3}};
    \node[square] (7) [fill=light-blue, above right of=6] {\textbf{7}};
    \node[square] (1) [fill=light-blue, above left of=4] {\textbf{1}};
    \node[square] (2) [fill=light-blue, below left of=4] {\textbf{2}};
    \node[main] (5) [below right of=10] {5};
    \node[main] (8) [above right of=9] {8};
    
    \draw[line width = 2pt, <-] (1) -- (2) node[midway, left] {5}; 
    \draw[-] (1) -- (3) node[midway, left] {6};
    \draw[-] (1) -- (4) node[midway, above, sloped] {100};
    \draw[-] (3) -- (6) node[midway, above, sloped] {7}; 
    \draw[line width = 2pt, <-] (6) -- (7) node[midway, above, sloped] {3};
    \draw[-] (6) -- (9) node[midway, above] {9};
    \draw[line width = 2pt, ->] (4) -- (10) node[midway, above] {2}; 
    \draw[line width = 2pt, ->] (10) -- (9) node[midway, right] {1};
    \draw[-] (2) -- (5) node[midway, above] {3};
    \draw[-] (5) -- (8) node[midway, right] {11};
    \draw[-] (9) -- (8) node[midway, above, sloped] {20};
    \draw[line width = 2pt, <-] (3) -- (7) node[midway, above] {3};
    \draw[-] (7) -- (8) node[midway, above] {1};
    \draw[line width = 2pt, ->] (2) -- (4) node[midway, above, sloped] {5};
    \end{tikzpicture}
    \caption{Optimal Solution using integer programming} \label{fig:fig4}
\end{minipage}
\end{figure*}

Different companies that provide cloud or edge computing services charge differently from the app vendors' hiring these resources. E.g., the cost of hosting applications in Amazon AWS depends on the individual user's usage, as it offers a pay-as-you-go approach. Similarly, the Oracle Cloud pricing model is based on a simple unit cost structure supporting a wide range of technology use cases. Thus, from an app vendor's perspective, the cost of the data transactions in an edge computing environment depends on the two components mentioned below \textemdash
\begin{itemize}
	\item cloud-to-edge (C2E) \textemdash \hspace{3pt}In this, the data transaction occurs between the cloud server and some of the edge servers called the \textit{initial transit edge server}\footnote{An initial transit edge server is an edge server that receives data directly from the cloud, and then distributes it to other edge servers.}.
	\item edge-to-edge (E2E) \textemdash \hspace{3pt}In this, the data transaction occurs between the edge servers.
\end{itemize}
To better estimate the total cost associated with the data transmission for the app vendor, we need to model an EDD solution strategy that comprises both these components, i.e., a C2E Strategy and an E2E Strategy as listed below \textemdash
\begin{itemize}
    \item C2E Strategy \textemdash \hspace{3pt}In order to handle the data transactions between the cloud and initial transit edge servers, we need to define a set $S$ = ($s$\textsubscript{1}, $s$\textsubscript{2} $\cdots$ $s$\textsubscript{N}), where $s\textsubscript{v}$ (1 $\leq$ $v$ $\leq$ $N$) denotes the set of boolean array representation of the initial transit edge servers, i.e., the nodes which receives data directly from the cloud.
    \begin{equation} \label{eq:constraint1}
        s\textsubscript{v} = 
        \begin{cases}
        1, & \text{if \textit{v} is an initial transit edge server}\\
        0, & \text{if \textit{v} is not an initial transit edge server}
        \end{cases}
    \end{equation}

    \item E2E Strategy \textemdash \hspace{3pt}In order to handle the data transactions between the edge servers, we need to define a set $T$ = ($\tau$\textsubscript{(1, 1)}, $\tau$\textsubscript{(1, 2)} $\cdots$ $\tau$\textsubscript{(N, N)}), where $\tau$\textsubscript{\textit{(u, v)}} (\textit{u, v} $\in$ \textit{V}) denotes whether the data is transmitted through edge \textit{e\textsubscript{(u, v)}} in the graph \textit{G}.
    \begin{equation} \label{eq:constraint2}
        \tau\textsubscript{\textit{(u, v)}} = 
        \begin{cases}
        1, & \text{if data is transmitted through {e\textsubscript{(u, v)}}}\\
        0, & \text{if data is not transmitted through {e\textsubscript{(u, v)}}}
        \end{cases}
    \end{equation}
\end{itemize}
As we mentioned before, the destination edge servers must be sent data from the cloud server, thus, any EDD solution strategy must connect each destination edge server \textit{v} $\in$ \textit{R} to an initial transit edge server in \textit{S} through edges in \textit{E}, i.e.,
\begin{equation} \label{eq:constraint3}
    \textit{Connected(S, T, u, v)} = True, \hspace{3pt} \forall \textit{v} \in \textit{R}, \hspace{3pt} \exists \hspace{2pt} u, s\textsubscript{u} = 1
\end{equation}

Now, we considered the length of the transmission links as a parameter for the delay constraint, as the propagation delay ($P\textsubscript{delay}$) is directly proportional to the link-length ($L\textsubscript{link}$), assuming the speed ($s\textsubscript{medium}$) in the medium to be constant, i.e.,
\begin{equation}
    P\textsubscript{delay} = \frac{L\textsubscript{link}}{s\textsubscript{medium}}
\end{equation}

We here assume that the bandwidth delay in the links and the queuing delay at the nodes is negligible, and thus the vendors' latency or delay constraint will depend just on the total length of the links. Also, since the mode of transmission between the cloud server and each of the edge server is same. Thus, we denote the propagation delay, or, corresponding length constraint between them as a constant, i.e., the weights of these edges in the graph $G(V, E, W)$ will be a constant denoted by $\gamma$.
\begin{equation}\label{eq:constraint-1}
    W\textsubscript{(c, v)} = \gamma, \forall \hspace{3pt} v \in V \setminus \{c\}
\end{equation}

Now, the app-vendor has the liberty to regulate the EDD delay constraint as per the application specifics and requirements. Let $P\textsubscript{limit}$ represent the app vendors' EDD propagation delay constraint, and $L\textsubscript{limit}$ represent the equivalent EDD length constraint. Thus, for each of the destination edge server $v$, the total path length between $v$ and its connected parent initial transit edge server in $S$, i.e., $L\textsubscript{v}$ must not exceed this EDD length constraint $L\textsubscript{limit}$.
\begin{equation} \label{eq:constraint4}
    0 \leq \textit{L\textsubscript{v}} \leq \textit{L\textsubscript{limit}}, \textit{L\textsubscript{v}} \in Z\textsuperscript{+}, \forall \textit{v} \in \textit{R}
\end{equation}

For example, consider the graph in \autoref{fig:fig3}, the nodes of the graph represent the edge servers given by
$V$ = \{1, 2, 3, 4, 5, 6, 7, 9\}, where number of edge servers $N$ = 9 and the set of transmission links i.e., $E$ = \{$e$\textsubscript{(1, 2)}, $e$\textsubscript{(1, 3)}, $e$\textsubscript{(1, 4)} $\cdots$ $e$\textsubscript{(7, 9)}\}. Now, let the EDD length constraint as per the vendor's application requirements be $L\textsubscript{limit}$ = 112 and $\gamma = 100$. This means each edge server must receive data within the maximum path length of 12, where path starts from an initial transit edge server. In \autoref{fig:fig3}, if node 4 is the only edge server selected as the initial transit edge server, then one possible E2E strategy is to select edges \{$e$\textsubscript{(1, 2)}, $e$\textsubscript{(1, 4)}, $e$\textsubscript{(1, 5)}, $e$\textsubscript{(2, 3)}, $e$\textsubscript{(4, 8)}\}. This means that in the set T\textsubscript{E2E} there is $\tau$\textsubscript{(1, 2)} =  $\tau$\textsubscript{(1, 4)} = $\tau$\textsubscript{(1, 5)} = $\tau$\textsubscript{(2, 3)} = $\tau$\textsubscript{(4, 8)} = 1. Accordingly, we can also obtain the depth of each of these destination edge servers, i.e, $L$\textsubscript{1} = 101, $L$\textsubscript{2} = 103, $L$\textsubscript{4} = 100, $L$\textsubscript{5} = 110, $L$\textsubscript{8} = 110, and $L$\textsubscript{3} = 110.

Now, given an EDD length constraint of $L\textsubscript{limit}$, the app vendor’s optimization objective is to minimize the EDD cost, which consists of the part incurred by the C2E transmissions and the E2E transmissions.

Thus, the Edge Data Distribution (EDD) problem can be formulated as \textemdash
\begin{equation}
    minimize \left( Cost\textsubscript{C2E}(S) + Cost\textsubscript{E2E}(T) \right)
\end{equation}
while fulfilling the EDD constraints in \autoref{eq:constraint3}, \autoref{eq:constraint-1}, and \autoref{eq:constraint4}.

As we can see, the Edge Data Distribution (EDD) is a constrained optimization problem based on a well-known NP-Hard problem in graph theory known as the Steiner Tree \cite{IEEEexample:papermy}. 
\section{Solution Strategy}
\hypertarget{SolSgy}{In} this section, we first introduce a modified implementation of an optimal approach based on \cite{IEEEexample:papermy}, to solve the EDD problem considering the end-to-end delay using the Integer Programming (IP) technique. The IP approach is exponential in time complexity and so will take an eternity to produce solution for the large datasets. Thus, we design an O(k) heuristic approach named EDD-NSTE to estimate the solutions for the large-scale EDD problems in polynomial time complexity. This will be followed by a theoretical analysis of the performance of the EDD-NSTE algorithm.
\subsection{Edge Data Distribution (EDD) as an Integer Programming Problem}
The Edge Data Distribution (EDD) can be formulated as a constrained integer programming problem as given in \cite{IEEEexample:papermy}. Our modified implementation of the EDD considering the end-to-end delay is specified as below.

The solution to the EDD problem must minimize the cost incurred for data transmissions while fulfilling the app vendor’s EDD length constraint of $L$\textsubscript{limit}. Thus, to model the EDD problem as a generalized constrained integer programming optimization problem firstly we add the cloud $c$ into $V$, and then add the edges from cloud $c$ to each edge server in graph $G$. Now, we can formulate the C2E strategy of EDD, $S$ = ($s$\textsubscript{1}, $s$\textsubscript{2}, $s$\textsubscript{3} $\cdots$ $s$\textsubscript{N}) by selecting edges in graph $G$, such that, T\textsubscript{c} = ($\tau$\textsubscript{(c, 1)}, $\tau$\textsubscript{(c, 2)} $\cdots$ $\tau$\textsubscript{(c, N)}), where $\tau$\textsubscript{(c, v)} denotes whether the data is transmitted from cloud server $c$ to the edge server $v$. Here, we combine the C2E and E2E strategy of EDD into one parameter
\begin{equation}
\begin{gathered}
    T = (\tau\textsubscript{(c, 1)}, \tau\textsubscript{(c, 2)} \cdots  \tau\textsubscript{(c, N)}, \tau\textsubscript{(1, 1)} \cdots  \tau\textsubscript{(N, N)}),\\
    where, \hspace{3pt} \tau\textsubscript{(u, v)} \in (0, 1) \hspace{3pt} (u \in V, v \in V \setminus\{c\})
\end{gathered}
\end{equation}
which indicates whether the edge $e$\textsubscript{(u, v)} is included in $T$. We also need to define a parameter for each of the edge servers, i.e.,
\begin{equation}
    H\textsubscript{v} = 
    \begin{cases}
    0, & \text{if v is not visited during the EDD process}\\
    1, & \text{if v is visited during the EDD process}
    \end{cases}
\end{equation}
here, $H\textsubscript{v} = 1, \forall v \in R$ will make sure that we include the destination edge servers into our solution. Also, $H$\textsubscript{$c$} = 1, as cloud will always be a part of the solution. Now, the constrained integer programming optimization problem can be formally expressed as follows, 
\begin{equation}\label{eq:eqMin}
    min \left( \mathop{\sum}_{v \hspace{2pt} \in \hspace{2pt} V \setminus \{c\}}\hspace{-5pt}\tau\textsubscript{(c, v)} \cdot W\textsubscript{(c, v)} \hspace{4pt} + \mathop{\sum\sum}_{u, v \hspace{2pt} \in \hspace{2pt} V \setminus \{c\}}\hspace{-5pt}\tau\textsubscript{(u, v)} \cdot W\textsubscript{(u, v)} \right)
\end{equation}
\begin{flushleft}
subjected to constraints,
\end{flushleft}
\begin{equation}
    H\textsubscript{v} = 1, \forall v \in R \cup \{c\}
\end{equation}
\begin{equation}\label{eq:const0}
    W\textsubscript{(c, v)} = \gamma, \forall \hspace{3pt} v \in V \setminus \{c\}
\end{equation}
\begin{equation}\label{eq:const1}
    \tau\textsubscript{(u, v)} \leq H\textsubscript{u}\cdot H\textsubscript{v}, \forall u \in V, v \in V \setminus \{c\}
\end{equation}
\begin{equation}\label{eq:const2}
    \sum \tau\textsubscript{(u, v)} = H\textsubscript{v}, \forall u \in V,  v \in V \setminus\{c\}
\end{equation}
\begin{equation}\label{eq:const3}
    \mathop{\sum_{v \in V \setminus \{c\}}\hspace{-9pt}\tau\textsubscript{(c, v)}} \geq 1
\end{equation}
\begin{equation}\label{eq:const4}
    \tau\textsubscript{(u, v)}, H\textsubscript{v} \in (0, 1)
\end{equation}
\begin{equation}\label{eq:const6}
    L\textsubscript{c} = 0, L\textsubscript{c} \leq L\textsubscript{v} \leq L\textsubscript{limit}, \forall v \in V \setminus \{c\}
\end{equation}
\begin{equation}\label{eq:const7}
    L\textsubscript{v} - L\textsubscript{u} = W\textsubscript{(u, v)}, \forall u, v \in V, \tau\textsubscript{(u, v)} = 1
\end{equation}
Here, in-order to minimize the total cost incurred by both the C2E and E2E edges given by \autoref{eq:eqMin}, we must satisfy the above constraints. The \autoref{eq:const0} makes sure that the weight of the edge from cloud to each of the edge server is equal to $\gamma$. The \autoref{eq:const1} will make sure that, if edge $e\textsubscript{(u, v)}$ is considered a part of the solution, or, $\tau\textsubscript{(u, v)} = 1$, then both the nodes $u$ and $v$ must be visited, i.e., $H\textsubscript{v} = 1$ and $H\textsubscript{u} = 1$. \autoref{eq:const2} suggests a constraint that the summation of $\tau\textsubscript{(u, v)}$ of the edges incoming at the particular node $v$, must be equal to the $H\textsubscript{v}$ of that particular node, i.e., whether or not the node $v$ is visited, if edges incoming onto it are considered a part of the solution. \autoref{eq:const3} suggests that the solution must have atleast one cloud server, i.e., summation of $\tau\textsubscript{(c, v)}$ must be greater than or equal to $1$. \autoref{eq:const6} suggests that the depth of any edge server $v$, must be within $[0, L\textsubscript{limit}]$, and \autoref{eq:const7} suggests that for nodes $u$, $v$ connected through edge $e\textsubscript{(u, v)}$ in the EDD solution must satisfy the depth difference of $W\textsubscript{(u, v)}$.

\subsection{Example to understand the Integer Programming Approach}
Consider an example in \autoref{fig:fig4} that displays the optimal solution generated, if we solve the graph in \autoref{fig:fig2} using the integer programming approach, with EDD length constraint of $L\textsubscript{limit}$ = 110, and $W\textsubscript{(c, v)} = 100, \forall v \in V \setminus \{c\}$. The C2E strategy will specify \textit{node 2} and \textit{node 7} as the initial transit edge servers which receives the data directly from the cloud server and thus, the cost incurred for the cloud to edge data transmission is \textit{Cost\textsubscript{C2E}} = 200. Now, these nodes will transmit the data to the other destination edge servers (represented by light blue square nodes) that can be reached within the given length constraint of \textit{L\textsubscript{limit}}. Thus, the cost incurred for the edge to edge data transmission is \textit{Cost\textsubscript{E2E} = 19}. The integer programming approach selects the edges E = \{e\textsubscript{(c, 2)}, e\textsubscript{(c, 7)}, e\textsubscript{(2, 1)}, e\textsubscript{(2, 4)}, e\textsubscript{(4, 10)}, e\textsubscript{(10, 9)}, e\textsubscript{(7, 3)}, e\textsubscript{(7, 6)}\}, for data transmissions and thus, the optimal total cost incurred is the sum of \textit{Cost\textsubscript{C2E}} and \textit{Cost\textsubscript{E2E}}, i.e., 219. 
\subsection{Edge Data Distribution (EDD) estimation as a Network Steiner Tree Estimation Problem}
\begin{algorithm}[tb!]
    \caption{\textbf{Network Steiner Tree Approximation}}
        F $\gets \overline{G}\textsubscript{R}$;\hspace{3pt}W $\gets$ $\phi$;\hspace{3pt}$Triples$ $\gets$ \{$z$ $\subset$ R :\norm{z}= 3\} \\
        \ForEach{z $\in$ $Triples$}{
            \textbf{find} v which \textbf{minimizes} the $\sum_{s \in z} d(v, s)$ \\
            v(z) $\gets$ v \\
            d(z) $\gets$ $\sum_{s \in z} d(v(z), s)$ \\
        }
        \While{true}{
            M $\gets$ an MST of F; $FindSave(M)$ \\
            \textbf{find} z $\in$ $Triples$ which \textbf{maximizes}, $win$ = $\max\limits_{e \in z}save(e)$ + $\min\limits_{e \in z}save(e)$ - d($z$)\\ 
            \If{win $\leq$ 0}{\textbf{break}}
            F $\gets$ F[z]; insert(W, v(z)) \\
        }
        Find a steiner tree for R $\cup$ W in graph G using MST algorithm
\end{algorithm}
\begin{algorithm}[b]
    \caption{\textbf{To calculate $FindSave(M)$}}
    \If{$M$ $\neq$ empty}{
        $e'$ $\gets$ an edge of $M$ with maximum weight\\ $x$ $\gets$ $d(e')$ \\
        $M$\textsubscript{1}, $M$\textsubscript{2} $\gets$ the subgraphs connected through $e'$ \\
        \ForEach{vertex $v$\textsubscript{1} of $M$\textsubscript{1} and $v$\textsubscript{2} of $M$\textsubscript{2}}{
            $save$($e$) $\gets$ $x$, where $e$ denotes $e\textsubscript{($v$\textsubscript{1}, $v$\textsubscript{2})}$ \\
        }
        $FindSave$($M$\textsubscript{1}); $FindSave$($M$\textsubscript{2})
    }
\end{algorithm}
\begin{algorithm}[t]
\caption{\textbf{EDD-NSTE Algorithm}}
\KwIn{G, L\textsubscript{limit}, pathLength, W, visited}
\KwOut{G\textsubscript{final}}
Construct the graph $G\textsubscript{ST}$ from \hyperlink{algo:algo1}{Algorithm 1}. Take the maximum connectivity node in the graph $G\textsubscript{ST}$ and connect it to the cloud to make it a rooted tree. Update the distances between consecutive nodes by the shortest path between them and let this graph be $G\textsubscript{DT}$\\
Initialize pathLength[c] $\gets$ 0, visited[c] $\gets$ 1, pathLength[v] $\gets$ $\infty$, visited[v] $\gets$ 0, $\forall$ v $\in$ V, Update G\textsubscript{DT} to include the minimum cost path between any two vertices with edges in graph G \\
Update G\textsubscript{DT} to form a directed rooted tree, with cloud as the root \\
Run a Breadth-First Search Algorithm to compute the minimum depth of each vertex v $\in$ G\textsubscript{DT} from cloud \\
\ForEach{vertex v, in the path from root to leaf node and parent p in Depth-First Search Algorithm}{
    \If{visited[v] $\neq$ 1}{
        \If{pathLength[v] $\leq$ L\textsubscript{limit} and v $\in$ R}{
            add edge e\textsubscript{(p, v)} in G\textsubscript{final} \\
            visited[v] $\gets$ 1
        }
        \ElseIf{v $\notin$ R}{
            add edge e\textsubscript{(p, v)} in G\textsubscript{final} \\
            visited[v] $\gets$ 1
        }
        \ElseIf{pathLength[v] $>$ L\textsubscript{limit} and v $\in$ R}{
            pathLength[v] $\gets$ W\textsubscript{(c, v)}, visited[v] $\gets$ 1\\
            add edge e\textsubscript{(c, v)} in G\textsubscript{final}\\
            \ForEach{child node u of node v in graph G in Depth-First Search order}{
                \If{pathLength[u] $>$ pathLength[v] + W\textsubscript{(u, v)} and pathLength[v] + W\textsubscript{(u, v)} $\leq$ L\textsubscript{limit}}{
                    \If{e(v, u) $\notin$ G\textsubscript{DT}}{
                        update edge in G\textsubscript{DT} from v to u\\
                        pathLength[u] $\gets$ pathLength[v] + W\textsubscript{(u, v)}
                    }
                }
            }
            run a Breadth-First Search Algorithm and update depths of the vertices in G\textsubscript{DT}
        }
    }
}
Run a Depth-First Search and remove the unnecessary edges from the graph G\textsubscript{final} \\
Calculate the total cost incurred as the sum of W\textsubscript{(u, v)} $\forall$ e\textsubscript{(u, v)} $\in$ G\textsubscript{final} 
\end{algorithm}
The Edge Data Distribution (EDD) is an NP-Hard \cite{IEEEexample:papermy} problem. Thus, finding an optimal solution for large-scale EDD scenarios is troublesome due to the exponential time complexity of the algorithm. To address this issue, we here present an estimation approach, named Edge Data Distribution as a Network Steiner Tree Estimation (EDD-NSTE), for finding the approximate solutions to large scale EDD problems efficiently. The approximation ratio for EDD-NSTE is $O(k)$, which means that the ratio of the EDD cost incurred by EDD-NSTE and that incurred by the optimal solution is $O(k)$ in the worst case, where $k$ is constant. 

We will here present an algorithm that calculates the estimated cost incurred for a given graph in two steps \textemdash
\begin{itemize}
    \item First of all, we calculate a network steiner tree approximation \cite{IEEEexample:paperZelikovsky} with destination edge servers $R$, as the set of distinguished vertices, in the given graph $G$.
    \item Then, we present a heuristic algorithm which estimates a rooted minimum steiner tree, by adding cloud $c$ to the graph, and then slicing and fine-tuning the rooted steiner tree with the vendor's EDD length constraint of $L\textsubscript{limit}$.
\end{itemize}
\begin{figure*}[t]
\centering
\begin{minipage}[b]{0.48\textwidth}
\centering
    \begin{tikzpicture}[node distance={16.5mm}, thick, main/.style = {draw, circle}, square/.style={regular polygon,regular polygon sides=4}] 
    \node[square] (6) [fill=light-blue] {\textbf{6}};
    \node[square] (9) [fill=light-blue,right of=6] {\textbf{9}};
    \node[main] (10) [below of=9] {10};
    \node[square] (4) [fill=light-blue, below of=6] {\textbf{4}};
    \node[square] (3) [fill=light-blue, above left of=6] {\textbf{3}};
    \node[square] (7) [fill=light-blue, above right of=6] {\textbf{7}};
    \node[square] (1) [fill=light-blue, above left of=4] {\textbf{1}};
    \node[square] (2) [fill=light-blue, below left of=4] {\textbf{2}};
    \node[main] (5) [below right of=10] {5};
    \node[main] (8) [above right of=9] {8};
    
    \draw[line width = 2pt, -] (1) -- (2) node[midway, left] {5}; 
    \draw[line width = 2pt, -] (1) -- (3) node[midway, left] {6};
    \draw[-] (1) -- (4) node[midway, above, sloped] {100};
    \draw[-] (3) -- (6) node[midway, above, sloped] {7}; 
    \draw[line width = 2pt, -] (6) -- (7) node[midway, above, sloped] {3};
    \draw[-] (6) -- (9) node[midway, above] {9};
    \draw[line width = 2pt, -] (4) -- (10) node[midway, above] {2}; 
    \draw[line width = 2pt, -] (10) -- (9) node[midway, right] {1};
    \draw[-] (2) -- (5) node[midway, above] {3};
    \draw[-] (5) -- (8) node[midway, right] {11};
    \draw[-] (9) -- (8) node[midway, above, sloped] {20};
    \draw[line width = 2pt, -] (3) -- (7) node[midway, above] {3};
    \draw[-] (7) -- (8) node[midway, above] {1};
    \draw[line width = 2pt, -] (2) -- (4) node[midway, above, sloped] {5};
    \end{tikzpicture}
    \caption{Steiner Tree estimated by Algorithm 1} \label{fig:ns1}
\end{minipage}
\begin{minipage}[b]{0.48\textwidth}
\centering
    \begin{tikzpicture}[node distance={16.5mm}, thick, main/.style = {draw, circle}, square/.style={regular polygon,regular polygon sides=4}] 
    \node[square] (6) [fill=light-blue] {\textbf{6}};
    \node[square] (9) [fill=light-blue,right of=6] {\textbf{9}};
    \node[main] (10) [below of=9] {10};
    \node[square] (4) [fill=light-blue, below of=6] {\textbf{4}};
    \node[square] (3) [fill=light-blue, above left of=6] {\textbf{3}};
    \node[square] (7) [fill=light-blue, above right of=6] {\textbf{7}};
    \node[square] (1) [fill=light-blue, above left of=4] {\textbf{1}};
    \node[square] (2) [fill=light-blue, below left of=4] {\textbf{2}};
    \node[main] (5) [below right of=10] {5};
    \node[main] (8) [above right of=9] {8};
    
    \draw[line width = 2pt, ->] (1) -- (2) node[midway, left] {5}; 
    \draw[line width = 2pt, ->] (1) -- (3) node[midway, left] {6};
    \draw[-] (1) -- (4) node[midway, above, sloped] {100};
    \draw[-] (3) -- (6) node[midway, above, sloped] {7}; 
    \draw[-] (6) -- (7) node[midway, above, sloped] {3};
    \draw[line width = 2pt, <-] (6) -- (9) node[midway, above] {9};
    \draw[-] (4) -- (10) node[midway, above] {2}; 
    \draw[-] (10) -- (9) node[midway, right] {1};
    \draw[-] (2) -- (5) node[midway, above] {3};
    \draw[-] (5) -- (8) node[midway, right] {11};
    \draw[-] (9) -- (8) node[midway, above, sloped] {20};
    \draw[line width = 2pt, ->] (3) -- (7) node[midway, above] {3};
    \draw[-] (7) -- (8) node[midway, above] {1};
    \draw[line width = 2pt, ->] (2) -- (4) node[midway, above, sloped] {5};
    \end{tikzpicture}
    \caption{Final EDD solution estimated by Algorithm 3} \label{fig:ns2}
\end{minipage}
\end{figure*}
Here, given a graph $G$, we first calculate the metric closure of the graph, denoted by $\overline{G}$, which is a complete graph of $G$, having edge lengths equal to the shortest distance between the vertices in $G$. $\overline{G}\textsubscript{R}$ denotes the subgraph of $\overline{G}$ induced by a vertex subset $R$ $\subseteq$ $V$, i.e., the subgraph on destination edge servers. We denote the minimum spanning tree (MST) of any graph $G$ as $mst(G)$. Now, given a metrically closed graph $G$, we may contract an edge $e$, i.e., reduce its length to 0, we represent such a resulting graph as $G[e]$. For any triple set of vertices, $z$ = ($u$, $v$, $w$) we represent $G[z]$ as a graph resulting from contraction of two edges i.e., $e$\textsubscript{(u, v)} and $e$\textsubscript{(v, w)}. $Triples$ denotes the set of all combination of 3 destination edge servers, $v(z)$ denotes the centroid of a particular triple, i.e., the vertices other than the destination edge servers whose sum of distance from a triple is minimum and $d(e)$ denotes the weight of the edge $e$. Our implementation of the algorithm proposed in \cite{IEEEexample:paperZelikovsky} is given in \hyperlink{algo:algo1}{Algorithm 1}, which loops continuously to find the best possible reduction in the cost of the minimum spanning tree of $F$, where $F = \overline{G}\textsubscript{R}$ initially. The algorithm does so by adding the three edges of $G$ having a common end node, i.e., the centroid, and consecutively removing the max-weight edge from each resulting cycle using the $save$ matrix, from the MST of $F$. This class of algorithms is also known as the Triple Loss Contracting Algorithm. The output of \hyperlink{algo:algo1}{Algorithm 1} will be an approximated steiner tree $G\textsubscript{ST}$, with vertices in R $\cup$ W and edges $e$ $\subset$ $\overline{E}$.

Here, we calculate a $save$ matrix for edges in the graph $F$, to estimate the maximum win for a given triple, where $save(v1, v2)$ is the minimum value, such that both ends $v1$ and $v2$ are in the same component of $F$, given by \autoref{eq:saveEq}. To calculate the $save$ matrix, a pseudo code for a recursive function $FindSave(M)$ is given in \hyperlink{algo:algo2}{Algorithm 2}. The implementation time complexity of this algorithm is $O(\norm{R}\textsuperscript{2})$.
\begin{equation} \label{eq:saveEq}
    save(v1, v2) = mst(F) - mst(F[e\textsubscript{($v1$, $v2$)}]), \forall \hspace{2pt} v1, v2 \in F
\end{equation}
Now, since we have an approximated steiner tree, we need to add cloud to the given graph by adding an edge from cloud to the maximum connectivity node in the approximated Steiner tree. The resulting graph will then be converted into a rooted tree, with cloud as the root. We do so by implementing a breadth-first search on $G\textsubscript{ST}$. We also need to convert the edges between any two nodes of $G\textsubscript{ST}$ by the shortest path between two nodes in graph $G$, as the steiner tree $G\textsubscript{ST}$, was approximated over the metric closure of graph $G$. Let this directed graph be $G\textsubscript{DT}$.

The algorithm for slicing and fine-tuning the graph to satisfy the vendor's EDD length constraint is presented in \hyperlink{algo:algo3}{Algorithm 3}. In this algorithm, we take the directed steiner tree $G\textsubscript{DT}$ and initialize the $pathLength[]$ and the $visited[]$ for each node of the graph, with $visited[c] = 1$, $pathLength[c] = 1$, and $pathLength[v] = \infty, \forall v \in V$. It then runs the Breadth-First Search to compute the minimum length of path of each node from the root node, assuming the cloud server as the root. After that, it runs the Depth-First Search and \textemdash
\begin{itemize}
	\item for each unvisited destination edge server with path length less than or equal to $L\textsubscript{limit}$, it adds the corresponding edge into the final solution $G\textsubscript{final}$.
	\item for each unvisited edge server, which is not a destination edge server, it again adds the corresponding edge into the final solution $G\textsubscript{final}$.
	\item for each unvisited destination edge server having a depth greater than $L\textsubscript{limit}$, it connects that edge server directly to the cloud and adds edge from cloud to this edge server in $G\textsubscript{final}$. It then runs a Depth-First Search to update the minimum depths of the remaining unvisited edge servers assuming this edge server as the root.
\end{itemize}
Finally, this algorithm removes the unnecessary edges from the graph $G\textsubscript{final}$ and calculates the cost of the solution as the sum of the cost of edges in $G\textsubscript{final}$.

\subsection{Example to understand the EDD-NSTE algorithm}
Consider the graph in the \autoref{fig:fig2}, with data transmission latency $L\textsubscript{limit} = 110$, and $W\textsubscript{(c, v)} = \gamma = 100, \forall v \in V \setminus \{c\}$. When we apply \hyperlink{algo:algo1}{2Algorithm 1} onto it, it selects edges \{e\textsubscript{(2, 1)}, e\textsubscript{(2, 4)}, e\textsubscript{(1, 3)}, e\textsubscript{(4, 10)}, e\textsubscript{(10, 9)}, e\textsubscript{(3, 7)}, e\textsubscript{(7, 6)}\} to construct the estimated steiner tree given in \autoref{fig:ns1}, using the triple loss contracting mechanism, which in this case is also the optimal steiner tree possible. Now, as we can see the \textit{node 1} is the node with the max-connectivity in the estimated steiner tree, thus, we add this node to the cloud directly, as given in \autoref{fig:ns2} to make a rooted EDD solution. The \textit{node 2}, \textit{node 3}, \textit{node 4}, and \textit{node 7} can be reached within the 110 unit distance from the cloud, thus edges \{e\textsubscript{(c, 1)}, e\textsubscript{(1, 2)}, e\textsubscript{(1, 3)}, e\textsubscript{(2, 4)}, e\textsubscript{(3, 7)}\} are included into the final EDD solution. When the algorithm reaches \textit{node 10}, it encounters \textit{node 9} exceeding the $L\textsubscript{limit}$. Now, since \textit{node 9} is encountered first, the algorithm adds \textit{node 9} to the cloud directly, which then updates the final EDD solution to include \textit{node 6} and \textit{node 9}, as they can be reached within data transmission constraint with \textit{node 9} as the root. The final EDD solution then becomes \{e\textsubscript{(c, 1)}, e\textsubscript{(c, 9)},  e\textsubscript{(1, 2)}, e\textsubscript{(1, 3)}, e\textsubscript{(2, 4)}, e\textsubscript{(3, 7)}, e\textsubscript{(9, 6)}\}, which incurs a total cost of 228 distance units, including both the C2E and E2E edges.

\subsection{Time Complexity analysis of EDD-NSTE algorithm}
There has been a wealth of research in approximating the best Steiner Tree for a given general graph as mentioned in \cite{IEEEexample:collectionpaperstudy}, with a lower bound of smaller than 1.01 \cite{IEEEexample:paerlowerbound} theoretically. In this paper, we first introduced an algorithm to calculate the Network Steiner Tree approximation, based on the algorithm proposed in \cite{IEEEexample:paperZelikovsky} which has a time complexity of $O(\norm{V}\norm{E} + R\textsuperscript{4})$ and an approximation ratio of $O(11/6)$. The same approximation ratio can be achieved with a faster implementation of $O(\norm{R}(\norm{E} + \norm{V}\norm{R} +  \norm{V}log\norm{V}))$ as mentioned in \cite{IEEEexample:paperZelikovskysecond}.

The algorithm EDD-NSTE has a time complexity of $O(\norm{V} + \norm{E})$ in the worst case. Thus, the total time complexity of the EDD-NSTE Algorithm along with network steiner tree estimation is $O((\norm{V} + \norm{E}) + \norm{V}\norm{E} + \norm{R}\textsuperscript{4})$ and in the worst case the time complexity becomes $O(\norm{V}\textsuperscript{4})$.

In the next subsection, we prove that the EDD-NSTE is an $O(k)$ approximation algorithm with experimental results suggesting that EDD-NSTE significantly outperforms the other three representative approaches in comparison with a performance margin of 80.35\%.
\subsection{EDD-NSTE is an O(k) approximation algorithm}
As discussed in the previous section, let tree $G\textsubscript{final}$ be the final EDD solution produced by \hyperlink{algo:algo3}{Algorithm 3}, $G\textsubscript{ST}$ be the minimum cost steiner tree approximated by \hyperlink{algo:algo1}{Algorithm 1}, and $G\textsubscript{cv}$ be the tree having cloud $c$ as the root and edges $e\textsubscript{(c, v)}$ , $\forall$ v $\in$ $G\textsubscript{final} \setminus \{c\}$. Also, let the cost associated with each of these trees be $Cost(G\textsubscript{final})$, $Cost(G\textsubscript{ST})$, and $Cost(G\textsubscript{cv})$ respectively.

Let $G\textsubscript{opt}$ be the optimal solution for the given EDD problem and the optimal cost associated be $Cost(G\textsubscript{opt})$, thus, as per the research \cite{IEEEexample:paperZelikovsky}, we can write
\begin{equation}
    Cost(G\textsubscript{ST}) = \frac{11}{6}\hspace{2pt}Cost(G\textsubscript{opt})
\end{equation}
Now, let $v\textsubscript{o} = c$ be the cloud server and $v\textsubscript{i}$ be the i-th edge server, where $i \in \{1, 2,...m\}$, that introduces additional paths during the DFS traversal. Also, let $K = L\textsubscript{limit} - \gamma$. Thus, for any edge server there exists two possibilities, i.e.,
\begin{itemize}
    \item $v\textsubscript{i}$ exceeds the EDD length constraint of $L\textsubscript{limit}$ and thus, we need to add an extra edge $(c, v\textsubscript{i})$, which in turn adds a cost of $Cost\textsubscript{(G\textsubscript{cv})}(c, v\textsubscript{i}) = \gamma$.
    \item $\exists$ a path $(c, v\textsubscript{i})$ as a path from $(c, v\textsubscript{i-1})$ and  $(v\textsubscript{i-1}, v\textsubscript{i})$, such that $v\textsubscript{i}$ does not exceed the EDD length constraint, which in turn adds a cost of $Cost\textsubscript{(G\textsubscript{final})}(c, v\textsubscript{i}) \leq Cost\textsubscript{(G\textsubscript{cv})}(c, v\textsubscript{i-1}) + Cost\textsubscript{(G\textsubscript{ST})}(v\textsubscript{i-1}, v\textsubscript{i})$.
\end{itemize}

\begin{figure*}[tb!]
\centering
\begin{minipage}[b]{0.48\textwidth}
\centering
\includegraphics[height=0.8\textwidth, width=1.1\textwidth]{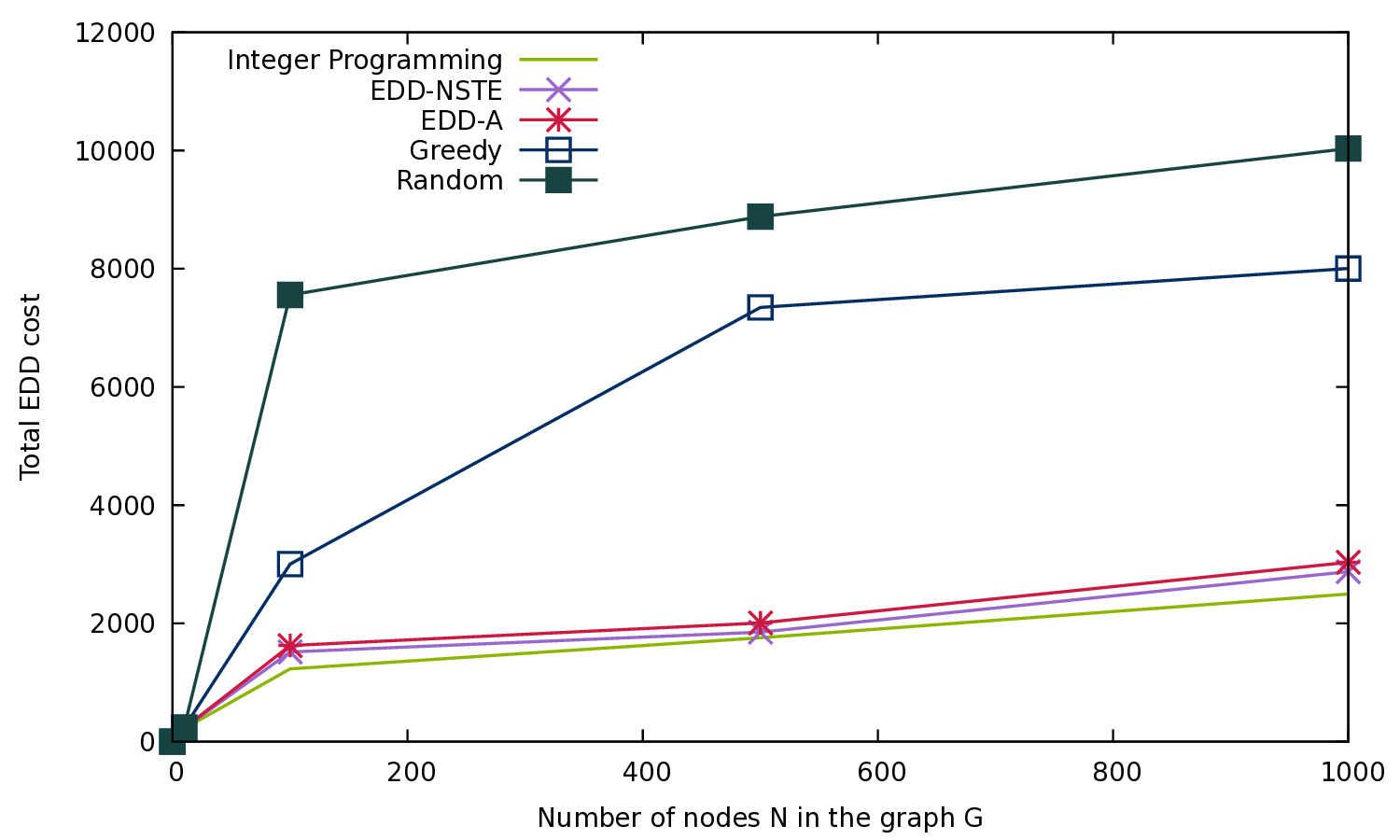}
\caption{Set 1: $N$ v/s Total EDD cost}\label{fig:fig5}
\end{minipage}\qquad
\hfill
\begin{minipage}[b]{0.48\textwidth}
\centering
\includegraphics[height=0.8\textwidth, width=1.1\textwidth]{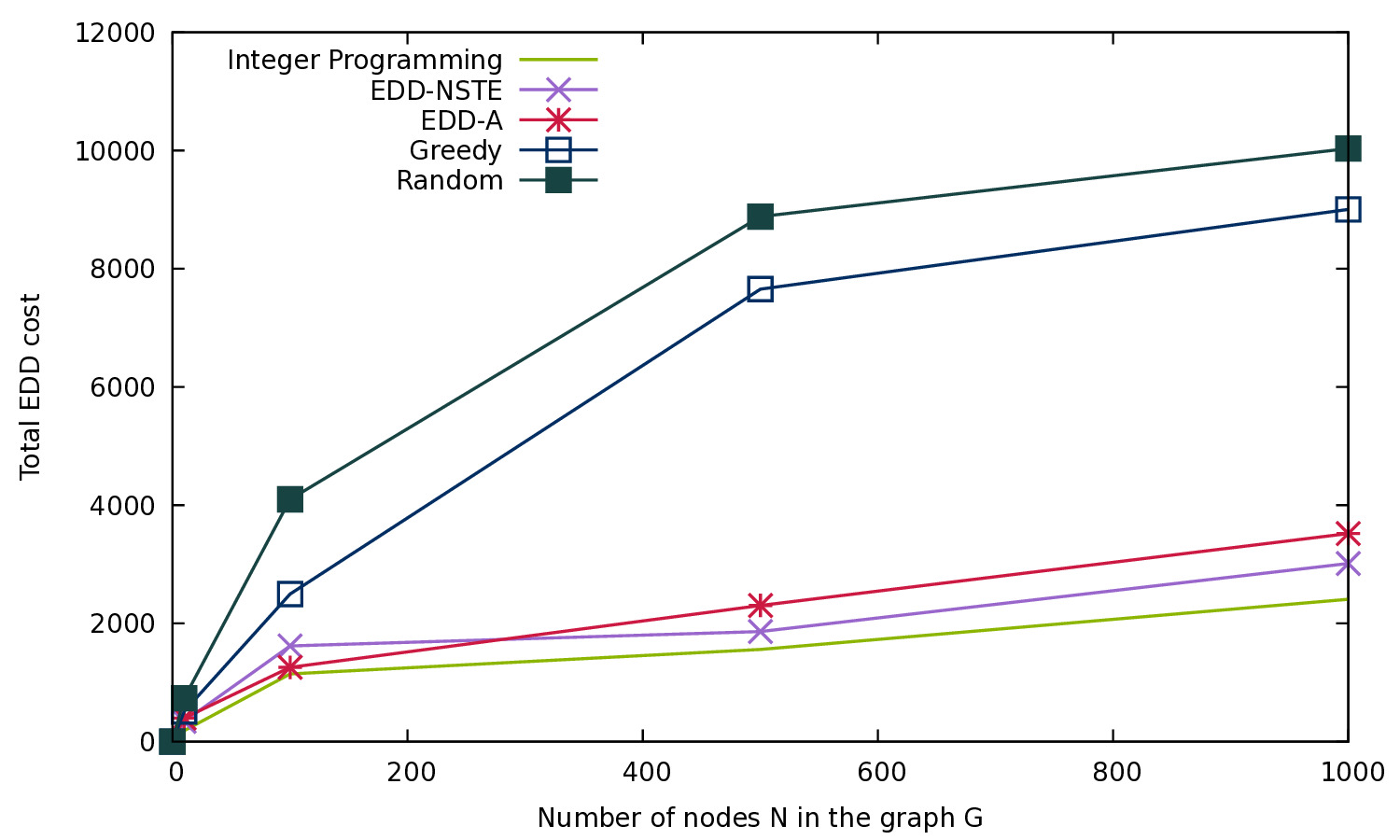}
\caption{Set 2: $N$ v/s Total EDD cost}\label{fig:fig25}
\end{minipage}
\end{figure*}
\begin{figure*}[tb!]
\centering
\begin{minipage}[b]{0.48\textwidth}
\centering
\includegraphics[height=0.8\textwidth, width=1.1\textwidth]{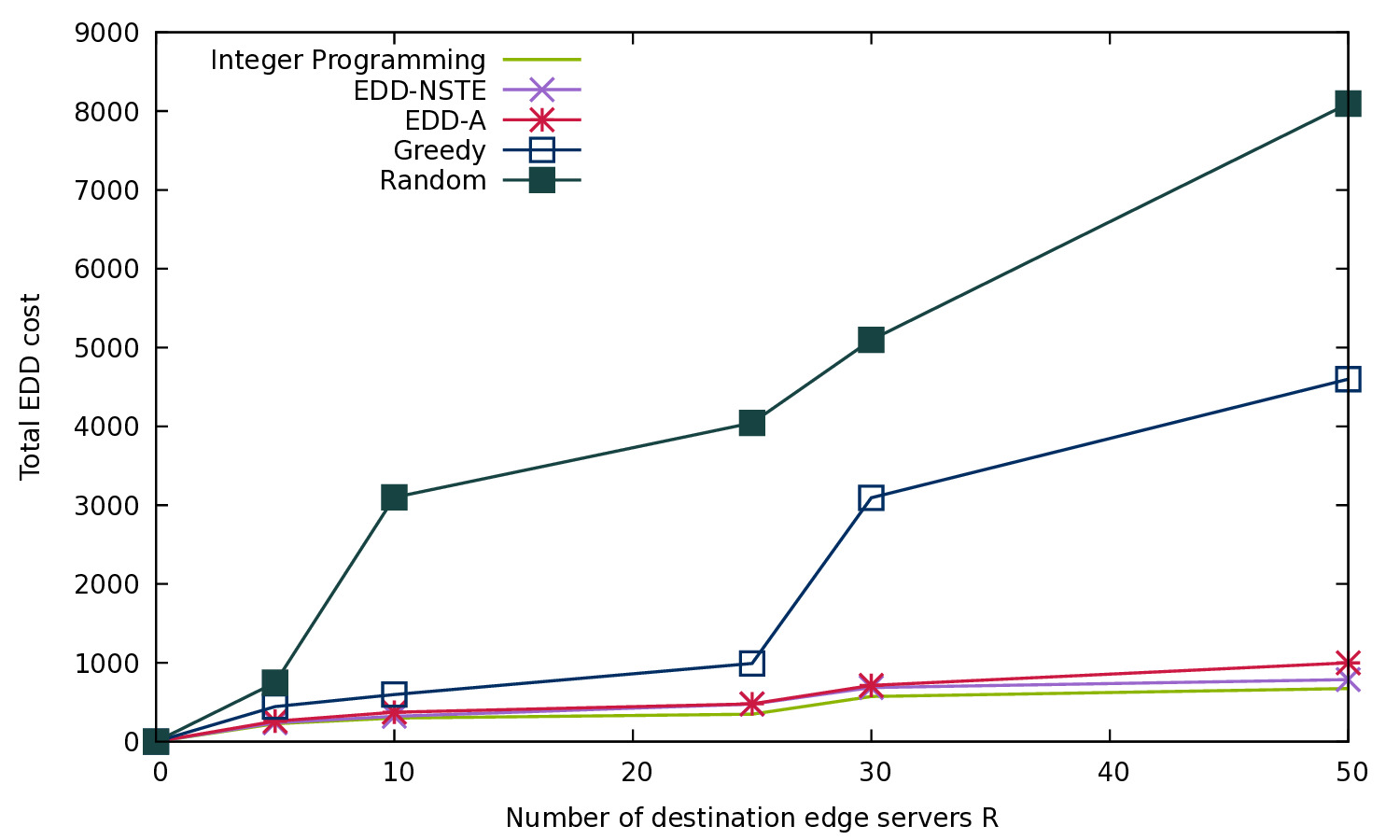}
\caption{Set 1: $R$ v/s Total EDD cost}\label{fig:fig6}
\end{minipage}\qquad
\hfill
\begin{minipage}[b]{0.48\textwidth}
\centering
\includegraphics[height=0.8\textwidth, width=1.1\textwidth]{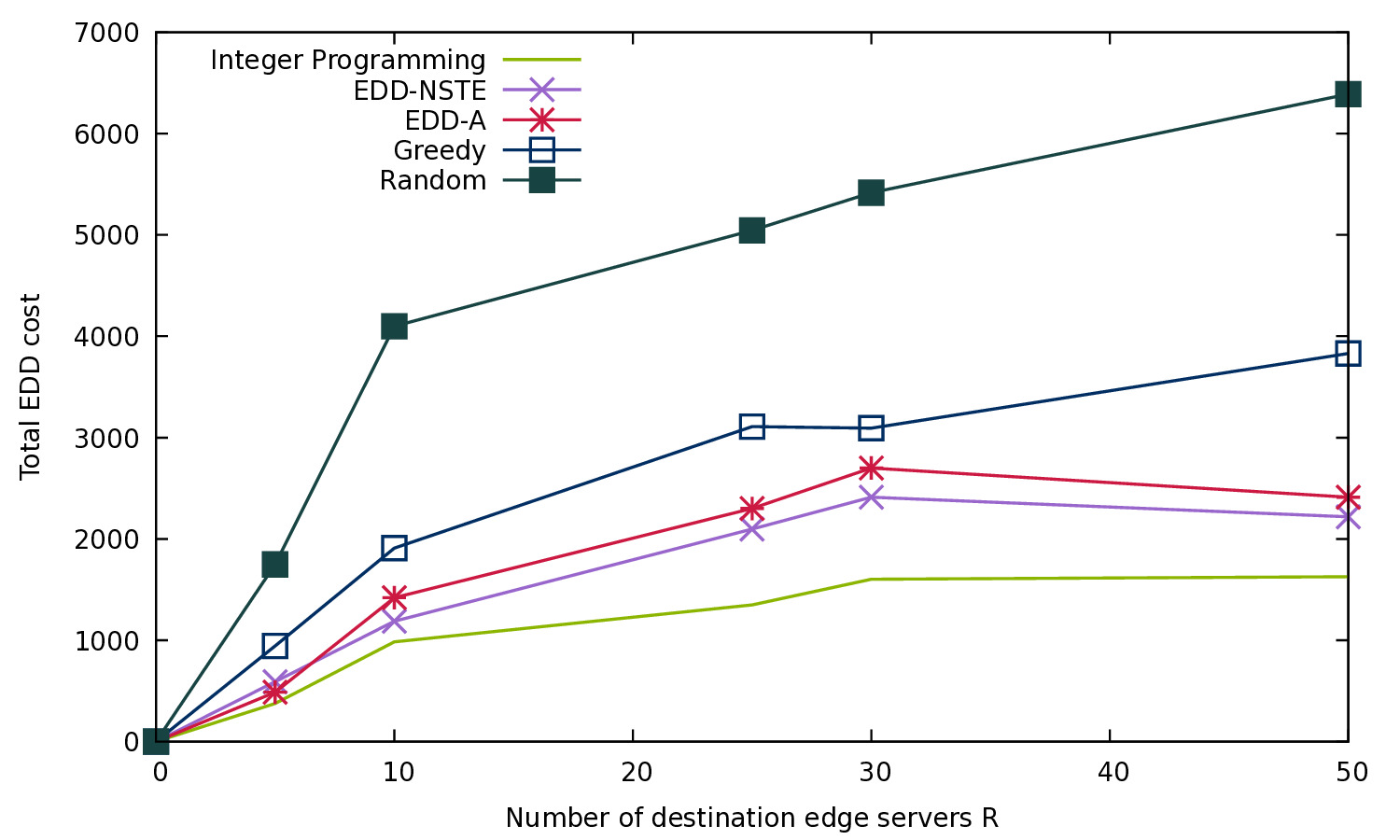}
\caption{Set 2: $R$ v/s Total EDD cost}\label{fig:fig26}
\end{minipage}
\end{figure*}
\begin{figure*}[tb!]
\centering
\begin{minipage}[b]{0.48\textwidth}
\centering
\includegraphics[width=0.77\textwidth, angle=-90]{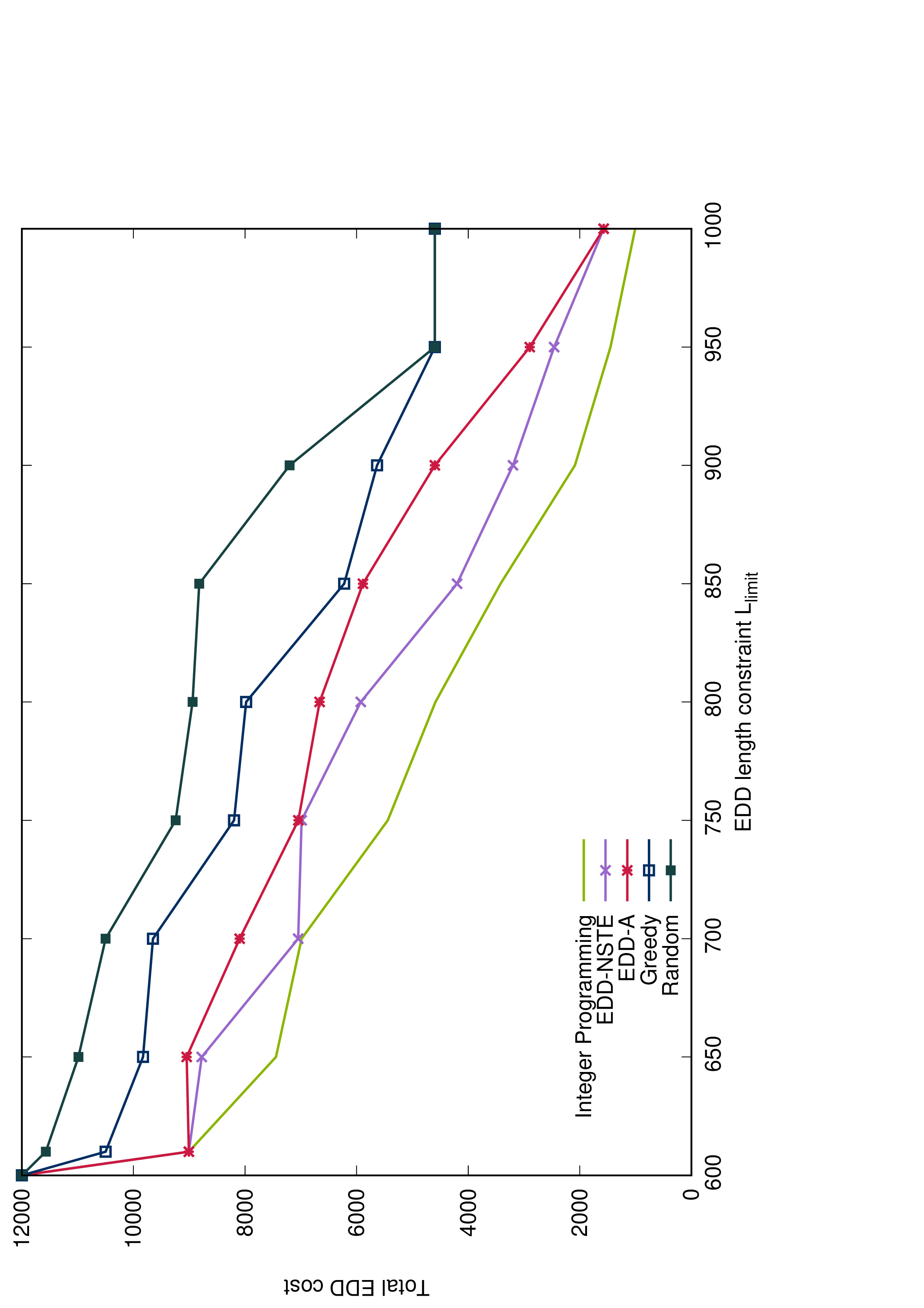}
\caption{Set 1: EDD length constraint $L\textsubscript{limit}$ v/s Total EDD cost}\label{fig:fig7}
\end{minipage}\qquad
\hfill
\begin{minipage}[b]{0.48\textwidth}
\centering
\includegraphics[width=0.77\textwidth, angle=-90]{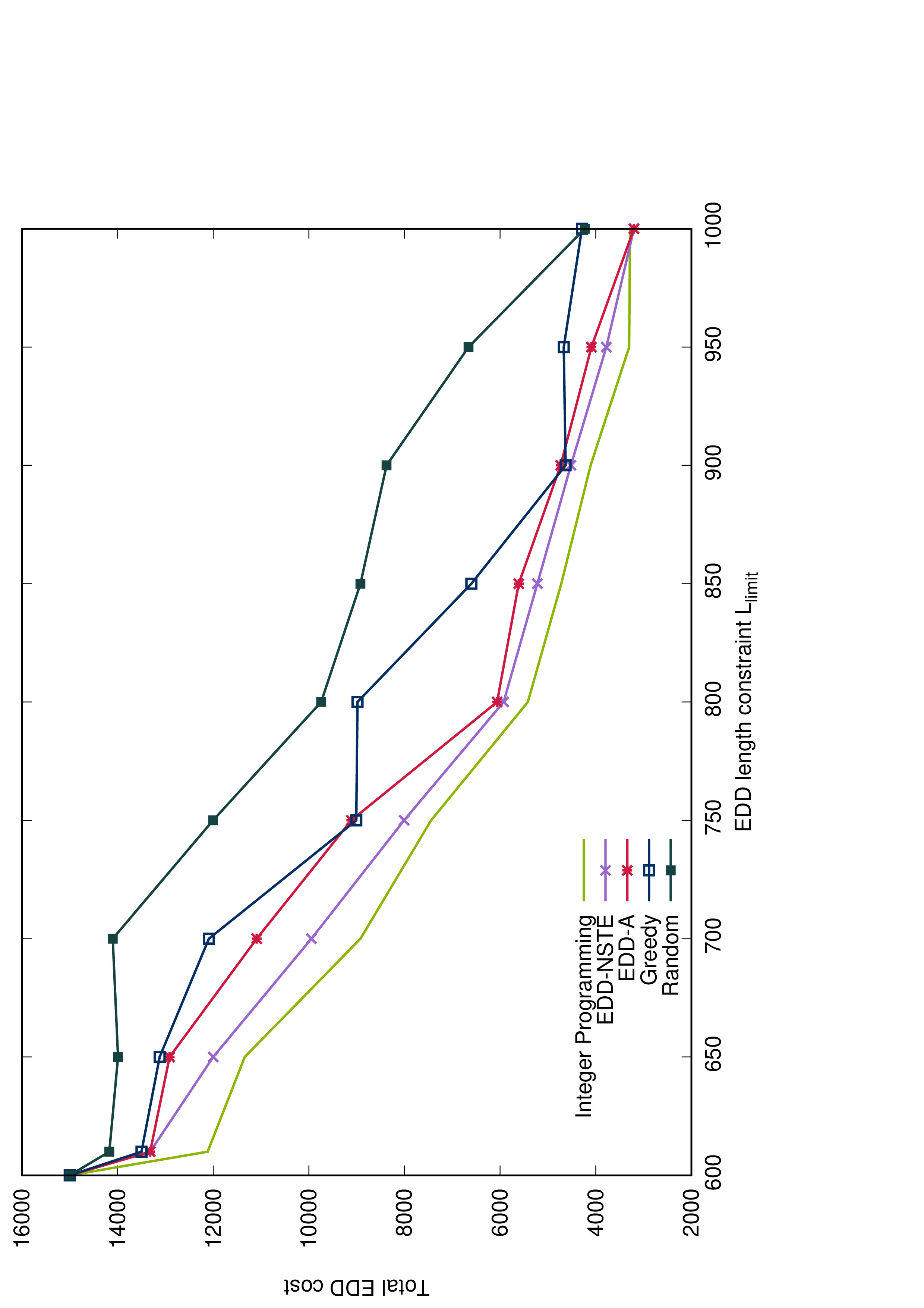}
\caption{Set 2: EDD length constraint $L\textsubscript{limit}$ v/s Total EDD cost}\label{fig:fig27}
\end{minipage}
\end{figure*}
\begin{figure*}[tb!]
\centering
\begin{minipage}[b]{0.48\textwidth}
\centering
\includegraphics[height=0.6\textwidth, width=1.1\textwidth]{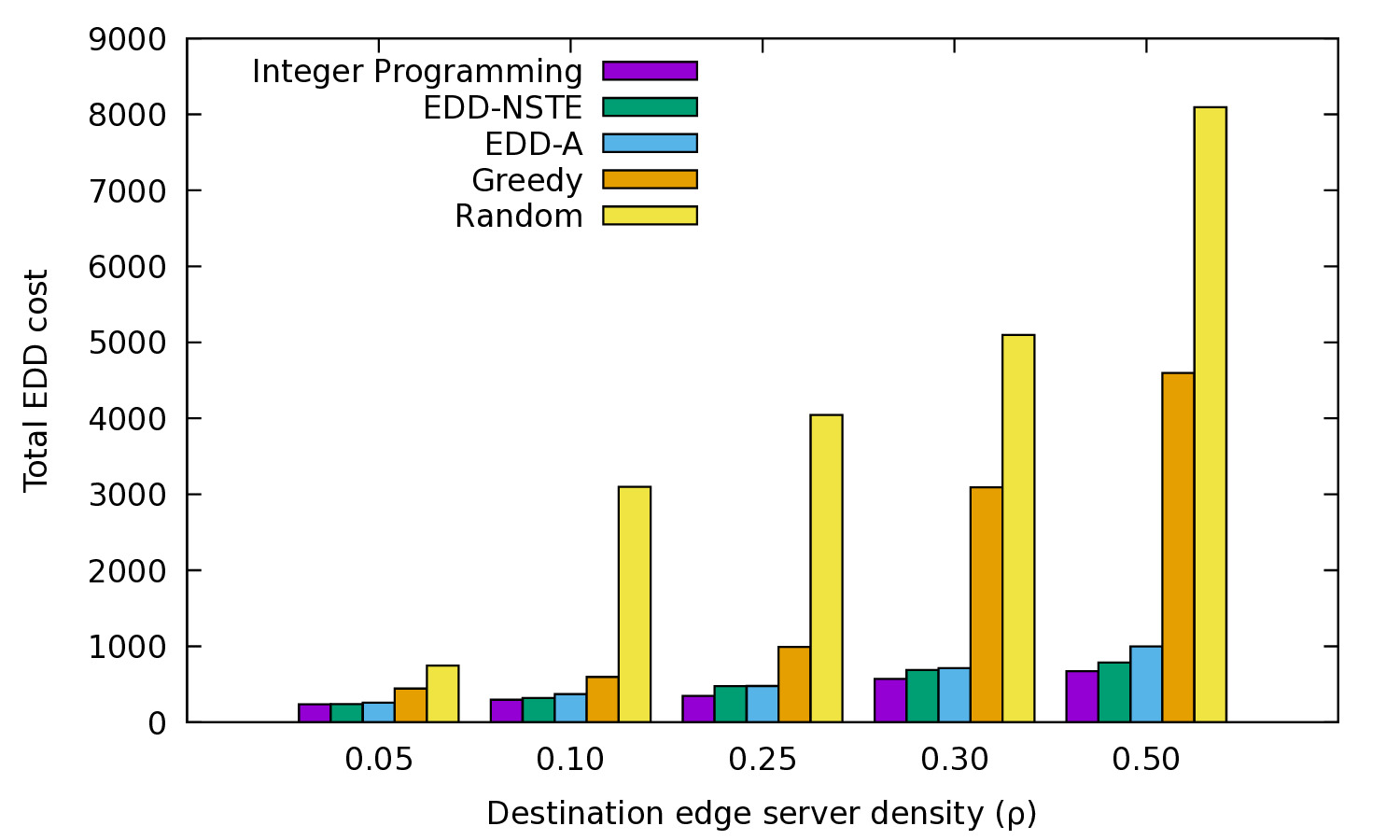}
\caption{Set 1: $\rho$ v/s Total EDD cost}\label{fig:fig8}
\end{minipage}\qquad
\hfill
\begin{minipage}[b]{0.48\textwidth}
\centering
\includegraphics[height=0.6\textwidth, width=1.1\textwidth]{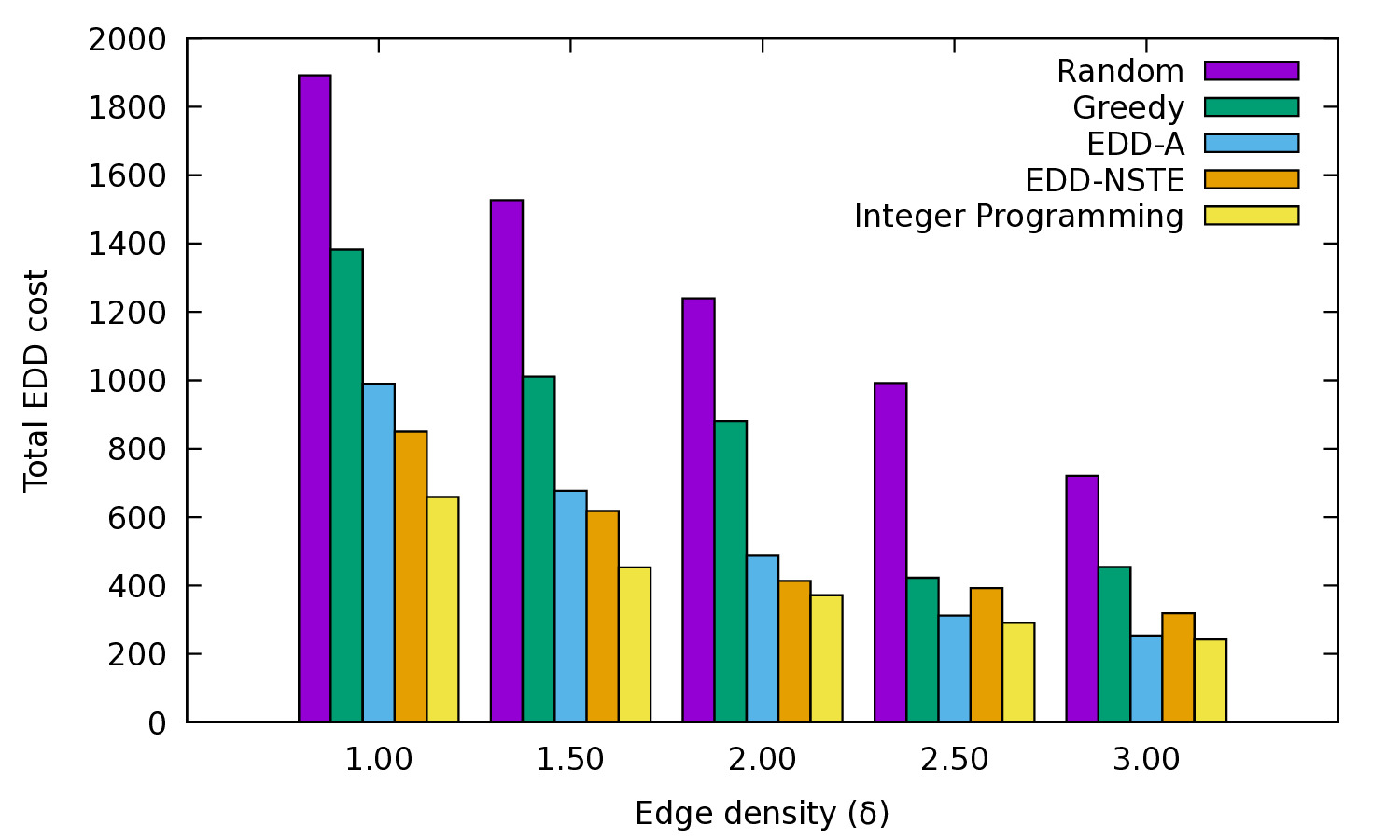}
\caption{Set 1: $\delta$ v/s Total EDD cost}\label{fig:fig9}
\end{minipage}
\end{figure*}   
\begin{figure*}[tb!]
\centering
\begin{minipage}[b]{0.48\textwidth}
\centering
\includegraphics[height=0.6\textwidth, width=1.1\textwidth]{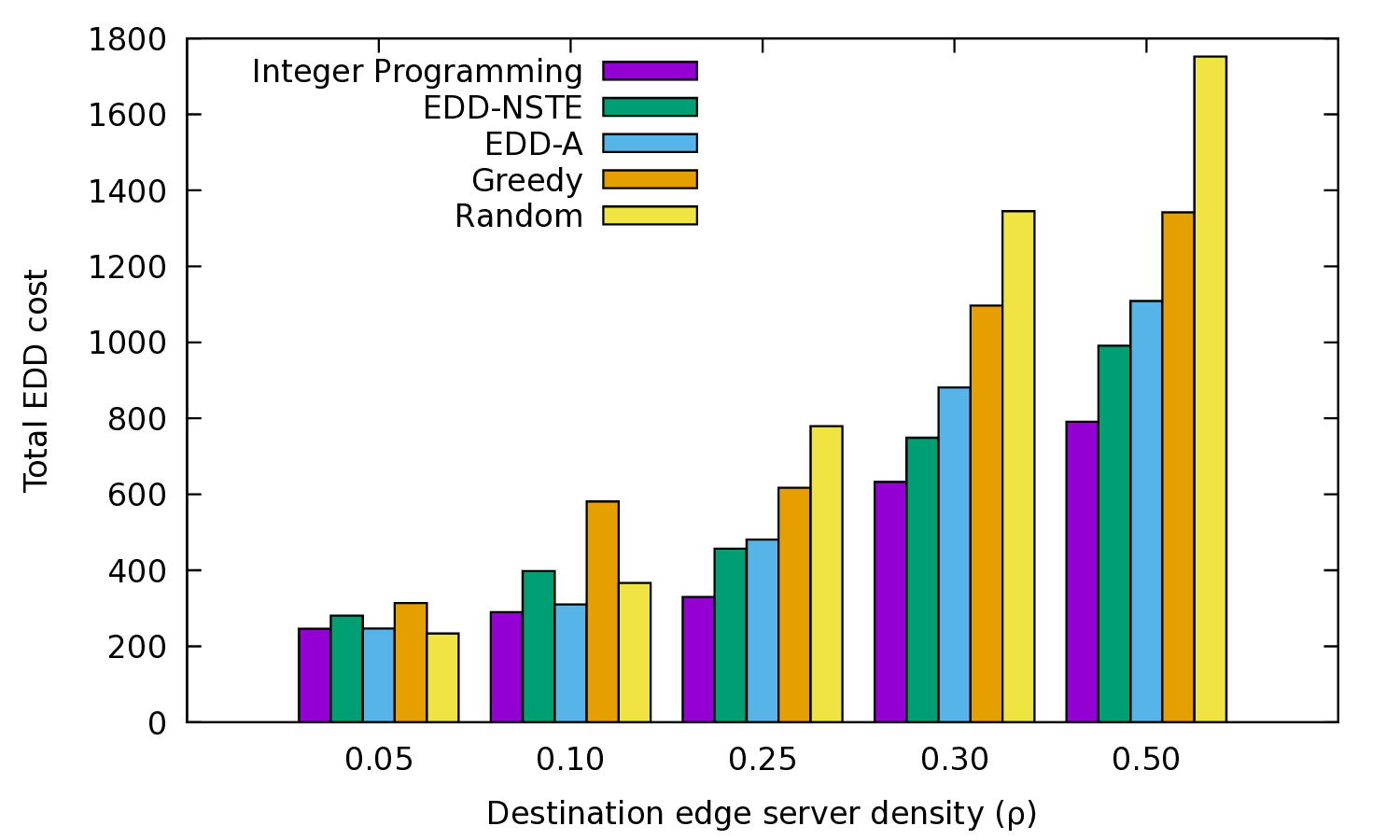}
\caption{Set 2: $\rho$ v/s Total EDD cost}\label{fig:fig28}
\end{minipage}\qquad
\hfill
\begin{minipage}[b]{0.48\textwidth}
\centering
\includegraphics[height=0.6\textwidth, width=1.1\textwidth]{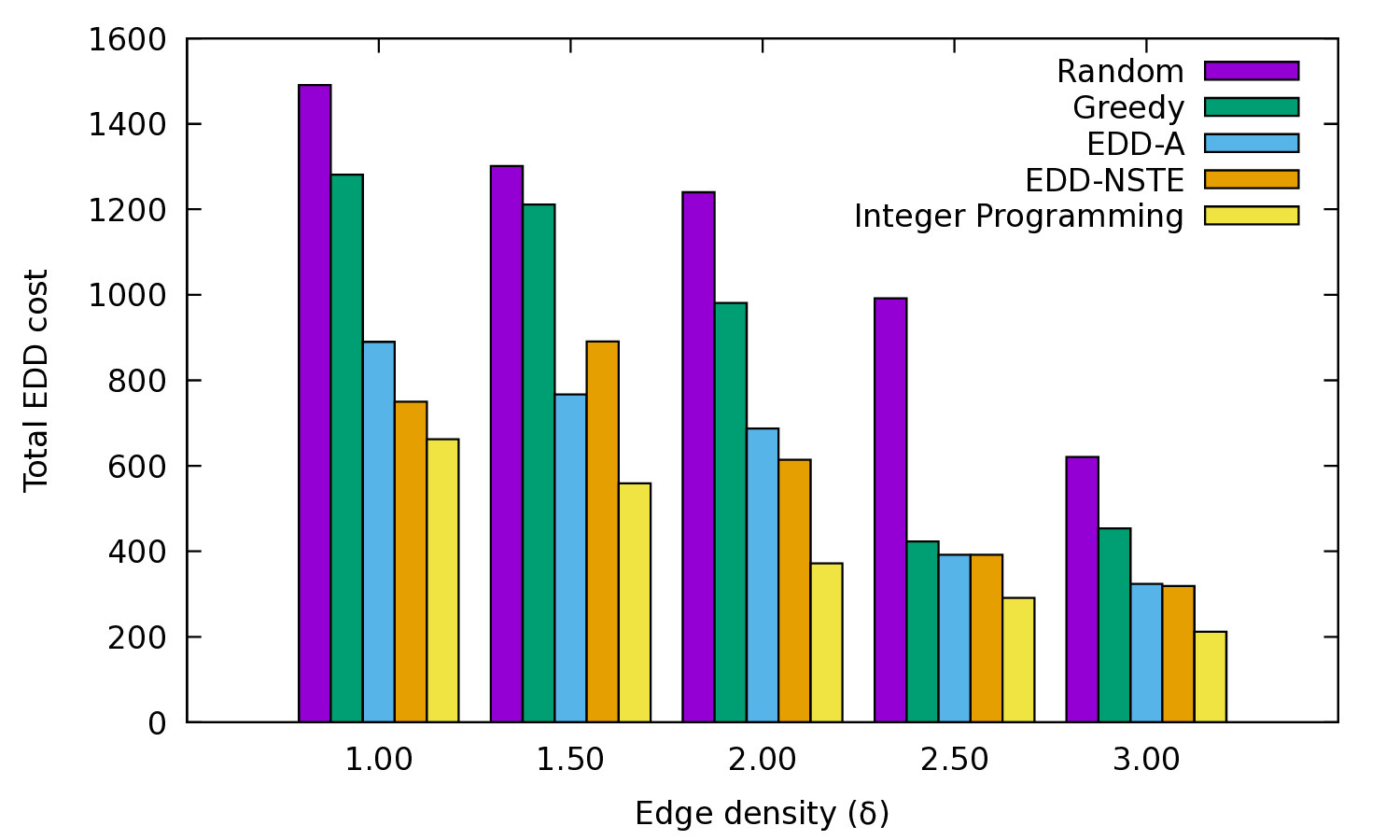}
\caption{Set 2: $\delta$ v/s Total EDD cost}\label{fig:fig29}
\end{minipage}
\end{figure*}
In the first case, when edge server $v\textsubscript{i}$ exceeds the EDD length constraint, then we must have
\begin{equation}
    Cost\textsubscript{(G\textsubscript{ST})}(c, v\textsubscript{i}) \geq \gamma + K \geq \left(1 + \frac{K}{\gamma}\right)Cost\textsubscript{(G\textsubscript{cv})}(c, v\textsubscript{i})
\end{equation}
where, $K$ denotes the parameter $L\textsubscript{limit} - \gamma$. Thus, we can obtain the combined equation as
\begin{equation}
\begin{gathered}
    \left(1 + \frac{K}{\gamma}\right)Cost\textsubscript{(G\textsubscript{cv})}(c, v\textsubscript{i}) \leq \\ Cost\textsubscript{(G\textsubscript{ST})}(c, v\textsubscript{i}) \leq Cost\textsubscript{(G\textsubscript{cv})}(c, v\textsubscript{i-1}) + Cost\textsubscript{(G\textsubscript{ST})}(v\textsubscript{i-1}, v\textsubscript{i})
\end{gathered}
\end{equation}
Summing the above equation for all of the $m$ edge servers, i.e., $v\textsubscript{i}$, where $i \in \{1, 2, ...m\}$, we have
\begin{equation}
\begin{gathered}
    \left(1 + \frac{K}{\gamma}\right) \sum_{i=1}^{m}Cost\textsubscript{(G\textsubscript{cv})}(c, v\textsubscript{i}) \leq \\ \sum_{i=1}^{m}Cost\textsubscript{(G\textsubscript{cv})}(c, v\textsubscript{i-1}) + \sum_{i=1}^{m}Cost\textsubscript{(G\textsubscript{ST})}(v\textsubscript{i-1}, v\textsubscript{i})
\end{gathered}
\end{equation}
Now, as we know from the basic inequality of positive numbers, that $\sum_{i=1}^{m-1}Cost\textsubscript{(G\textsubscript{cv})}(c, v\textsubscript{i}) \leq \sum_{i=1}^{m}Cost\textsubscript{(G\textsubscript{cv})}(c, v\textsubscript{i})$, so the above equation yields \textemdash
\begin{equation}
    \frac{K}{\gamma}\sum_{i=1}^{m}Cost\textsubscript{(G\textsubscript{cv})}(c, v\textsubscript{i}) \leq \sum_{i=1}^{m}Cost\textsubscript{(G\textsubscript{ST})}(v\textsubscript{i-1}, v\textsubscript{i})
\end{equation}
\begin{equation}
    \frac{K}{\gamma}Cost(G\textsubscript{cv}) \leq \sum_{i=1}^{m}Cost\textsubscript{(G\textsubscript{ST})}(v\textsubscript{i-1}, v\textsubscript{i})
\end{equation}
For each edge server $v\textsubscript{i}$, that changes its path in $G\textsubscript{final}$ during the DFS traversal without adding the path $(c, v\textsubscript{i})$, the update cost must be less than that of $Cost\textsubscript{(G\textsubscript{cv})}(c, v\textsubscript{i})$ and thus, the total cost after constructing $G\textsubscript{ST}$ cannot be more than $\sum_{i=1}^{m}Cost\textsubscript{(G\textsubscript{cv})}(c, v\textsubscript{i})$. Now, each edge in $G\textsubscript{ST}$ is traversed at most twice during the DFS traversal, thus, we have
\begin{equation}
    \sum_{i=1}^{m}Cost\textsubscript{(G\textsubscript{ST})}(v\textsubscript{i-1}, v\textsubscript{i}) \leq 2 \cdot Cost(G\textsubscript{ST})    
\end{equation}
Thus, combining both above equations, results in 
\begin{equation}
    Cost(G\textsubscript{cv}) \leq \frac{2\gamma}{K} Cost(G\textsubscript{ST})
\end{equation}
Finally, as we know, that the cost of the final EDD solution approximated by EDD-NSTE is
\begin{equation}
    Cost(G\textsubscript{final}) \leq Cost(G\textsubscript{cv}) + Cost(G\textsubscript{ST})
\end{equation}
\begin{equation}
    Cost(G\textsubscript{final}) \leq \frac{2\gamma}{K} Cost(G\textsubscript{ST}) + Cost(G\textsubscript{ST})
\end{equation}
\begin{equation}
    Cost(G\textsubscript{final}) \leq \left(\frac{2\gamma}{K} + 1\right) Cost(G\textsubscript{ST})
\end{equation}
\begin{equation}
    Cost(G\textsubscript{final}) \leq \frac{11}{6} \left(\frac{2\gamma}{K} + 1\right) Cost(G\textsubscript{opt})
\end{equation}
Now, since $K$ and $\gamma$ are both constants, we can safely assume $k = \frac{11}{6} \left(\frac{2\gamma}{K} + 1\right)$, and thus, EDD-NSTE is an $O(k)$ approximation algorithm.
\begin{figure*}[t]
    \centering
    \begin{minipage}[b]{0.48\textwidth}
\centering
    \begin{tikzpicture}[node distance={30mm}, thick, main/.style = {draw, circle}, square/.style={regular polygon,regular polygon sides=4}] 
    \node[square] (1) [fill=light-blue] {\textbf{1}};
    \node[square] (2) [fill=light-blue,right of=1] {\textbf{2}};
    \node[square] (3) [fill=light-blue,below of=2] {\textbf{3}};
    
    \draw[line width = 2pt, -] (1) -- (2) node[midway, above] {2}; 
    \draw[line width = 2pt, -] (2) -- (3) node[midway, right] {3};
    \end{tikzpicture}
    \caption{EDD-A considering hop distances} \label{fig:edda1}
\end{minipage}
\begin{minipage}[b]{0.48\textwidth}
\centering
    \begin{tikzpicture}[node distance={15mm}, thick, main/.style = {draw, circle}, square/.style={regular polygon,regular polygon sides=4}] 
    \node[square] (1) [fill=light-blue] {\textbf{1}};
     \node[main] (4) [right of=1] {};
    \node[square] (2) [fill=light-blue,right of=4] {\textbf{2}};
    \node[main] (5) [below of=2] {};
    \node[main] (6) [below of=5] {};
    \node[square] (3) [fill=light-blue,below of=6] {\textbf{3}};
    \draw[line width = 2pt, -] (1) -- (4) node[midway, above] {1}; 
    \draw[line width = 2pt, -] (4) -- (2) node[midway, above] {1};
    \draw[line width = 2pt, -] (2) -- (5) node[midway, right] {1};
    \draw[line width = 2pt, -] (5) -- (6) node[midway, right] {1};
    \draw[line width = 2pt, -] (6) -- (3) node[midway, right] {1};
    
    \end{tikzpicture}
    \caption{EDD-A considering propagation delays} \label{fig:edda2}
\end{minipage}
\end{figure*}

\section {Experimental Evaluation}
We experimentally evaluated the performance of above methods with the other simulation settings and approaches in comparison as discussed below. All experimental were conducted on an Ubuntu 20.04.1 LTS machine equipped with Intel Core i5-7200U processor (4 CPU’s, 2.50GHz) and 16GB RAM. The integer programming simulation was done on Google Collab (Python 3 Google Computer Engine Backend, 12.72GB RAM).
\subsection{Simulation Settings and Approaches for comparison of the result}
We compared our experimental results with the other optimization techniques and representative approaches \textemdash
\begin{itemize}
    \item \textbf{Greedy Connectivity (GC)} \textemdash \hspace{2pt} In this approach, for each node, the approach define its connectivity which is equal to the number of destination servers that have not received the data yet and are reachable within the vendors' EDD length constraint. Then it select the node with the maximum connectivity and make it an initial transmit edge server which then transmits the data to other servers within the app vendors' EDD length limit of \textit{L\textsubscript{limit}} until all destination edge servers have received the data.
    \item \textbf{Random} \textemdash \hspace{2pt} In this approach, it randomly select the initial transit edge servers which directly receive data from the cloud and then transmit it to the other servers within the app vendors' EDD length limit of \textit{L\textsubscript{limit}}, until all destination edge servers have received the data.
    \item \textbf{EDD Integer Programming} \textemdash \hspace{2pt} As discussed above in the \hyperlink{SolSgy}{Section IV}, we refine the formulation of given EDD problem into a 0-1 integer programming problem which can then be solved by using any simple IP solvers\footnote{IP solvers are the frameworks or tools used to solve integer programming problems. For E.g., \href{https://www.gnu.org/software/glpk/}{GLPK}, \href{http://lpsolve.sourceforge.net/5.5/}{LP Solve}, \href{https://developers.google.com/optimization}{Google OR Tools}, \href{https://pypi.org/project/PuLP/}{PuLP}, etc.}. This method take a huge amount of time to produce the result in the case of large datasets as is exponential in complexity.  
    \item \textbf{EDD-A Algorithm} \textemdash \hspace{2pt} This is state of the art approach given in \cite{IEEEexample:papermy}, in this approach it first calculate a connectivity-oriented minimum Steiner tree (CMST) using an $O(2)$ approximation method. This method calculates a minimum Steiner tree having cost at most twice the optimal Steiner tree on the graph. It then uses a heuristic algorithm to calculate the minimum cost of transmission incurred using E2E and C2E edges in the CMST. This algorithm is proved to have an $O(k)$ approximation solution and a time complexity $O(\norm{V}\textsuperscript{3})$, in the worst case.
    
    To compare this algorithm in the propagation delay scenario we have introduced temporary nodes in the graph which corresponds to the 1-hop distance in the edge-weight graph, i.e., \autoref{fig:edda1} represents the graph with 2 nodes and edge weights. In the scenario of hop distances these edge-weights can simply be taken as 1, but in the case of propagation delay we need to convert this graph into \autoref{fig:edda2} where temporary nodes between the two nodes will be same as the 1-hop distance edge length. Thus, now we can compare the results against EDD-A with vendor's latency limit (in terms of hop-distance), i.e., $D\textsubscript{limit} = L\textsubscript{limit}$.
\end{itemize}
\begin{figure*}[tb!]
\centering
\begin{minipage}[b]{0.22\textwidth}
\centering
\begin{tikzpicture}
    \begin{axis}[
            xlabel={$N$},
            ylabel={Time (in s)},
            xmin=0, xmax=100,
            ymin=0, ymax=10,
            xtick={0, 50, 100},
            ytick={},
            width=\textwidth,
            height=\textwidth,
            legend pos=north west,
            ymajorgrids=true,
            grid style=dashed,
            mark size = 1pt,
            every axis plot/.append style={line width=0.5pt},
        ]
        \addplot[
            color=eddste,
            every mark/.append style={solid, fill=eddste}, mark=square*,
            ]
            coordinates {
            (0,0)(10,0.5)(25,0.60)(50,1)(70,1.9)(85,5)(100,10)
            };
        \end{axis}
\end{tikzpicture}
\caption{$N$ v/s Computational Overhead}\label{fig:fig10}
\end{minipage}\qquad
\hfill
\begin{minipage}[b]{0.22\textwidth}
\centering
\begin{tikzpicture}
    \begin{axis}[
            xlabel={$\rho$},
            ylabel={Time (in s)},
            xmin=0, xmax=2.0,
            ymin=0, ymax=10,
            xtick={0.0, 0.5, 1.0, 1.5, 2.0},
            ytick={},
            width=\textwidth,
            height=\textwidth,
            legend pos=north west,
            ymajorgrids=true,
            grid style=dashed,
            mark size = 1pt,
            every axis plot/.append style={line width=0.5pt},
        ]
        \addplot[
            color=eddste,
            every mark/.append style={solid, fill=eddste}, mark=square*,
            ]
            coordinates {
            (0,0)(0.5,0.6)(1.0,1.5)(1.5,2.9)(1.8,4.6)(2.0,8.9)
            };
        \end{axis}
\end{tikzpicture}
\caption{$\rho$ v/s Computational Overhead}\label{fig:fig11}
\end{minipage}\qquad
\hfill
\begin{minipage}[b]{0.22\textwidth}
\centering
\begin{tikzpicture}
    \begin{axis}[
            xlabel={$L\textsubscript{limit}$},
            ylabel={Time (in s)},
            xmin=0, xmax=500,
            ymin=0, ymax=10,
            xtick={},
            ytick={},
            width=\textwidth,
            height=\textwidth,
            legend pos=north west,
            ymajorgrids=true,
            grid style=dashed,
            mark size = 1pt,
            every axis plot/.append style={line width=0.5pt},
        ]
        \addplot[
            color=eddste,
            every mark/.append style={solid, fill=eddste}, mark=square*,
            ]
            coordinates {
            (0,0)(100,1.8)(200,2.6)(300,6.2)(400,5.4)(500,3.4)
            };
        \end{axis}
\end{tikzpicture}
\caption{$L\textsubscript{limit}$ v/s Computational Overhead}\label{fig:fig12}
\end{minipage}\qquad
\hfill
\begin{minipage}[b]{0.22\textwidth}
\centering
\begin{tikzpicture}
    \begin{axis}[
            xlabel={$\delta$},
            ylabel={Time (in s)},
            xmin=0, xmax=2.0,
            ymin=0, ymax=100,
            xtick={0.0, 0.5, 1.0, 1.5, 2.0},
            ytick={},
            width=\textwidth,
            height=\textwidth,
            legend pos=north west,
            ymajorgrids=true,
            grid style=dashed,
            mark size = 1pt,
            every axis plot/.append style={line width=0.5pt},
        ]
        \addplot[
            color=eddste,
            every mark/.append style={solid, fill=eddste}, mark=square*,
            ]
            coordinates {
            (0,0)(0.5,20)(1.0,30)(1.5,50)(2.0,100)
            };
        \end{axis}
\end{tikzpicture}
\caption{$\delta$ v/s Computational Overhead}\label{fig:fig13}
\end{minipage}
\end{figure*}
\subsection{Dataset Used}
The experiments were conducted on a widely used EUA\footnote{https://github.com/swinedge/eua-dataset} and SLNDC\footnote{https://snap.stanford.edu/data/} Dataset. The EUA dataset contains the geographical locations of 1,464 real-world base stations in Melbourne, Australia. The set of destination edge servers \textit{R} is generated randomly. The links between the edge-servers are also generated randomly using an online random connected graph model generator tool. In the experiments, $\gamma$ = 100 is taken.
\subsection{Experiment Results}
Different EDD scenarios and algorithms mentioned above in the section \hyperlink{SolSgy}{solution strategy} and other approaches are simulated and compared to each other by varying different parameters as mentioned below \textemdash
\begin{itemize}
    \item The number of nodes $N$ in the graph $G$.
    \item The number of destination edge servers $R$ in the graph.
    \item The vendor's EDD length limit, i.e., $L\textsubscript{limit}$.
\end{itemize}
Among the above three factors, we vary one of the factors while keeping the others constant and then compare the estimated total EDD cost incurred in each case with the optimal cost generated by the integer programming approach. The experiment is done in two sets \textemdash
\begin{itemize}
	\item Set 1 \textemdash \hspace{3pt} Above factors are varied and the results are calculated on EUA Dataset.
	\item Set 2 \textemdash \hspace{3pt} Above factors are varied and the results are calculated on SLNDC Dataset.
\end{itemize}
\autoref{fig:fig5} and \autoref{fig:fig25}, displays the relationship between the number of nodes $N$ in the graph $G$ v/s total EDD cost incurred by different simulation settings and approaches. The value of the other parameters such as $L\textsubscript{limit} = 550$, $R = 25$ random nodes, $\gamma = 500$, and weights of the edges are in the range [1, 50] were fixed during the experiment for both the sets. As we can see that for both sets of experiments EDD-A and EDD-NSTE perform better than any other heuristic approach. For $N < 100$ there is a very slight difference between other approaches but as we increase the value of $N$, the difference becomes significantly more, and overall EDD-NSTE performs better than other heuristic methods. 

Similarly, \autoref{fig:fig6} and \autoref{fig:fig26}, displays the relationship between the number of destination edge servers $R$ v/s the total EDD cost incurred by different other settings. For this experiment the value of other parameters such as $N = 100$, $L\textsubscript{limit} = 250$, $\gamma = 200$, and weight of the edges are in the range [1, 50] were fixed during the experiment. As we can see that in Set 1, EDD-NSTE and EDD-A perform very close to the optimal solution, and in Set 2, there is a considerable difference between these two heuristics.

Now, the only factor left to alter is $L\textsubscript{limit}$, and the rest of the parameters are kept constant similar to the previous case. The number of destination edge servers $\norm{R} = 20$ for Set 1, and $\norm{R} = 25$ for Set 2, $\gamma = 600$ in this experiment. \autoref{fig:fig7} and \autoref{fig:fig27}, displays the relationship between the vendor's EDD length constraint $L\textsubscript{limit}$ v/s the total EDD cost incurred by different simulations. As we know, these two quantities are inversely related, i.e., with an increase in the value of EDD length constraint, the total EDD cost decreases. The total EDD cost reaches its maxima of $\gamma$R when $L\textsubscript{limit}$ = $\gamma$, as in this case all the destination edge servers are connected directly to the cloud server to fulfill the vendors' EDD length constraint. Also, as we can see that the Random and Greedy Connectivity algorithms become constant for $N \geq 950$ in Set 1, as then almost any random or greedy selection of servers is not able to improve on the EDD cost.

Here, we consider another parameter referred to as the destination edge server density. It is defined as the ratio of number of destination edge server upon total number of edge servers, i.e., Destination Edge Server Density ($\rho$) \textemdash
\begin{equation}
    \rho = \frac{\norm{R}}{\norm{V}}
\end{equation}
\autoref{fig:fig8} and \autoref{fig:fig28}, displays the bar-graph relationship between the total EDD cost and the density of the destination edge servers. As we can see, with the increase in density, the total EDD cost associated with each of the simulations tends to increase as a whole. Among the other algorithms and approaches in comparison, EDD-NSTE Algorithm performed well overall.

Similarly, we consider another experimental parameter known as edge density. This parameter helps to estimate the overall density of the graph. A very dense graph will have a higher value of $\delta$ than a sparse graph or a tree. Thus, it is defined as the total no of edges or transmission links in the graph upon total no of edge servers, i.e, Edge Density ($\delta$) \textemdash
\begin{equation}
    \delta = \frac{\norm{E}}{\norm{V}}
\end{equation}
\autoref{fig:fig9} and \autoref{fig:fig29}, displays the bar-graph relationship between the total EDD cost incurred and the density of the graph, or edge density. As we can see, with the increase in the edge density, the total EDD cost associated with each of the simulations tends to decrease as now shorter paths from cloud to other edge servers exist and thus, any destination edge server can be reached within the vendors' EDD length constraint following these shorter paths, thereby decreasing the number of initial transit edge servers, which in turn decreases the total EDD cost incurred as the $Cost\textsubscript{C2E}$ is significantly greater than $Cost\textsubscript{E2E}$.

\autoref{fig:fig10} to \autoref{fig:fig13} displays the computational overhead of EDD-NSTE algorithm versus different other graph parameters like number of nodes($N$), destination edge server density($\rho$), the vendors' EDD length constraint($L\textsubscript{limit}$), and edge density($\delta$) respectively. As we can see, with an increase in $N$, $\delta$ or $\rho$, the computational overhead of EDD-NSTE increases, as its worst-case time complexity is $O(\norm{V}\textsuperscript{4})$, whereas, with an increase in $L\textsubscript{limit}$, the computational overhead first increases and then decreases, as relaxing the vendors' EDD length constraint after a certain limit decreases the extra overhead caused by the slicing and pruning algorithm.

\section{Conclusion}
The EDD-NSTE algorithm suggested in this paper performs considerably better than the other simulations and approaches implemented for the edge data distribution problem. The total EDD cost approximated by this algorithm is closer to the optimal solution than the other representative approaches like Greedy Connectivity or Random. The algorithm also performs significantly better than the modified version of the last best EDD-A algorithm suggested in \cite{IEEEexample:papermy}. The EDD-A algorithm is an $O(k)$ approximation algorithm and the EDD-NSTE algorithm is an $O(k')$ approximation algorithm where $k' < k$ and thus, the performance of EDD-NSTE is better than EDD-A, which coincides with the experimental evaluations done so far.

EDD-NSTE algorithm produced better solutions than EDD-A or other simulations in more than 80.35\% of the test cases, ranging from very dense graphs to sparse graphs or trees.

Greedy Connectivity may produce solutions better than EDD-A or EDD-NSTE in some of the cases, but the greedy solutions are just an ad-hoc solution and thereby cannot always guarantee to produce better results whereas, the latter algorithms are restricted by their approximation ratio to always produce solutions below a certain maximum value for a given graph thus, they perform better than these ad-hoc methods on an average.
\bibliographystyle{./bibtex/bib/IEEEtran}
\bibliography{./bibtex/bib/IEEEabrv,./bibtex/bib/IEEEexample}

\begin{thebibliography}{10}
\providecommand{\url}[1]{#1}
\csname url@samestyle\endcsname
\providecommand{\newblock}{\relax}
\providecommand{\bibinfo}[2]{#2}
\providecommand{\BIBentrySTDinterwordspacing}{\spaceskip=0pt\relax}
\providecommand{\BIBentryALTinterwordstretchfactor}{4}
\providecommand{\BIBentryALTinterwordspacing}{\spaceskip=\fontdimen2\font plus
\BIBentryALTinterwordstretchfactor\fontdimen3\font minus
  \fontdimen4\font\relax}
\providecommand{\BIBforeignlanguage}[2]{{%
\expandafter\ifx\csname l@#1\endcsname\relax
\typeout{** WARNING: IEEEtran.bst: No hyphenation pattern has been}%
\typeout{** loaded for the language `#1'. Using the pattern for}%
\typeout{** the default language instead.}%
\else
\language=\csname l@#1\endcsname
\fi
#2}}
\providecommand{\BIBdecl}{\relax}
\BIBdecl

\bibitem{IEEE-CC-ACM}
M.~Armbrust and et~al., ``{A view of cloud computing},'' \emph{Commun. ACM},
  vol.~53, no.~4, pp. 50--58, 2010.

\bibitem{rfc8793}
B.~Wissingh, C.~A. Wood, and et~al., ``Information-centric networking {(ICN):}
  content-centric networking (ccnx) and named data networking {(NDN)}
  terminology,'' \emph{{RFC}}, vol. 8793, pp. 1--17, 2020.

\bibitem{Pathan2008}
M.~Pathan and R.~Buyya, \emph{{A Taxonomy of CDNs}}.\hskip 1em plus 0.5em minus
  0.4em\relax Berlin, Heidelberg: Springer Berlin Heidelberg, 2008, pp. 33--77.

\bibitem{GARVR}
``{https://arvr.google.com/}.''

\bibitem{GoogleARVR}
J.~Ren, Y.~He, G.~Huang, G.~Yu, Y.~Cai, and Z.~Zhang, ``An edge-computing based
  architecture for mobile augmented reality,'' \emph{{IEEE Network}}, vol.~33,
  pp. 162--169, 2019.

\bibitem{IEEEexample:papermy}
H.~Xia, F.~Chen, Q.~He, J.~C. Grundy, M.~Abdelrazek, and H.~Jin,
  ``{Cost-Effective App Data Distribution in Edge Computing},'' \emph{IEEE
  Tran. on Parallel and Distributed Systems}, vol.~32, no.~1, pp. 31--43, 2021.

\bibitem{IEEEexample:paperpiyush}
D.~Zhao, M.~Mohamed, and H.~Ludwig, ``Locality-aware scheduling for containers
  in cloud computing,'' \emph{IEEE Transactions on Cloud Computing}, vol.~8,
  no.~2, pp. 635--646, 2018.

\bibitem{IEEEexample:paperA}
H.~Yao, C.~Bai, M.~Xiong, D.~Zeng, and Z.~Fu, ``Heterogeneous cloudlet
  deployment and user-cloudlet association toward cost effective fog
  computing,'' \emph{Concurrency Comp. Pract. Exp.}, vol.~29, no.~16, 2017,.

\bibitem{IEEEexample:paperB}
H.~Yin, X.~Zhang, H.~H. Liu, Y.~Luo, C.~Tian, S.~Zhao, and F.~Li, ``Edge
  provisioning with flexible server placement,'' \emph{IEEE Trans. on Parallel
  Distribution System}, vol.~28, no.~4, p. 1031–1045, 2017.

\bibitem{IEEEexample:paperF}
P.~Lai, Q.~He, M.~Abdelrazek, F.~Chen, and et~al., ``Optimal edge user
  allocation in edge computing with variable sized vector bin packing,'' in
  \emph{Proc. Int. Conf. Service-Oriented Comput.}, pp. 230-245, 2018, Dec.
  5--9, 2018, pp. 230--245.

\bibitem{IEEEexample:papersparsh}
P.~Lai, Q.~He, J.~Grundy, and et~al., ``Cost-effective app user allocation in
  an edge computing environment,'' \emph{IEEE Transactions on Cloud Computing},
  vol.~31, no.~15, pp. 1--13, 2020.

\bibitem{IEEEexample:paperC}
X.~Cao, J.~Zhang, and H.~V. Poor, ``An optimal auction mechanism for mobile
  edge caching,'' in \emph{Proc. 38th IEEE Int. Conf. Distrib. Comput. Syst.},
  pp. 388-399, 2018, Dec. 5--9, 2018, pp. 388--399.

\bibitem{IEEEexample:paperD}
U.~Drolia, K.~Guo, J.~Tan, R.~Gandhi, and P.~Narasimhan, ``{Cachier:
  Edge-caching for recognition applications},'' in \emph{Proc. 37th IEEE Int.
  Conf. Distrib. Comput. Syst.}, pp. 276-286, 2017, Dec. 5--9, 2017, p.
  276–286.

\bibitem{IEEEExample:paperX}
K.~Zhang, S.~Leng, Y.~He, S.~Maharjan, and Y.~Zhang, ``{Cooperative Content
  Caching in 5G Networks with Mobile Edge Computing},'' \emph{IEEE Wireless
  Communications}, vol.~25, no.~3, pp. 80--87, 2018.

\bibitem{IEEEExample:paperY}
R.~Halalai, P.~Felber, A.-M. Kermarrec, and F.~Taïani, ``Agar: A caching
  system for erasure-coded data,'' in \emph{Proc. 37th IEEE Int. Conf. Distrib.
  Comput. Syst.}, pp. 23-33, 2017, Dec. 5--9, 2017, pp. 23--33.

\bibitem{IEEEExample:paperZ}
X.~Zhang and Q.~Zhu, ``{Hierarchical caching for statistical QoS guaranteed
  multimedia transmissions over 5G edge computing mobile wireless networks},''
  \emph{IEEE Wireless Comm.}, vol.~25, no.~3, pp. 12--20, 2018.

\bibitem{IEEEexample:paperE}
M.~Breitbach, D.~Schäfer, J.~Edinger, and C.~Becker, ``Context-aware data and
  task placement in edge computing environments,'' in \emph{Proc. IEEE Int.
  Conf. Pervasive Comput. Commun.}, pp. 1-10, 2019, Dec. 5--9, 2019, pp. 1--10.

\bibitem{IEEEexample:paperZelikovsky}
A.~Z. Zelikovsky, ``An 11/6-approximation algorithm for the network steiner
  problem,'' \emph{Algorithmica}, vol.~9, no.~5, pp. 463--470, 1993.

\bibitem{IEEEexample:collectionpaperstudy}
C.~Gröpl, S.~Hougardy, T.~Nierhoff, and J.~Prömel, ``Approximation algorithms
  for the steiner tree problems in graphs,'' \emph{Springer Verlag Berlin
  Heidelberg}, vol.~46, no.~5, pp. 235--279, 2001.

\bibitem{IEEEexample:paerlowerbound}
C.~Gröpl, S.~Hougardy, T.~Nierhoff, and H.~J. Prömel, ``Lower bounds for
  approximation algorithms for the steiner tree problem,'' \emph{Springer
  Verlag Berlin Heidelberg}, vol.~46, no.~5, pp. 217--228, 2001.

\bibitem{IEEEexample:paperZelikovskysecond}
A.~Z. Zelikovsky, ``A faster approximation algorithm for the steiner tree
  problem in graphs,'' \emph{Information Processing Letters}, vol.~46, no.~5,
  pp. 79--83, 1993.

\end{thebibliography}

\end{document}